\documentclass[%
 reprint,
superscriptaddress,
 amsmath,amssymb,
 aps,
floatfix,
]{revtex4-2}

\usepackage{graphicx}
\usepackage{dcolumn}
\usepackage{bm}
\usepackage{hyperref}

\usepackage{subcaption} 
\usepackage{ragged2e} 
\usepackage{booktabs} 
\usepackage{multirow} 
\usepackage[version=4]{mhchem} 
\usepackage{siunitx} 
\usepackage[capitalise]{cleveref}

\hypersetup{
  colorlinks   = true, 
  urlcolor     = blue, 
  linkcolor    = blue, 
  citecolor   = blue 
}

\begin{document}


\title{Cesium Clustering and Fluoroberyllate Network Disruption in FLiBe: A Total Scattering and Molecular Dynamics Study}

\author{Sean Fayfar}
\affiliation{Nuclear Reactor Laboratory, Massachusetts Institute of Technology, 77 Massachusetts Ave., Cambridge, MA 02139, USA}

\author{Rajni Chahal}
\affiliation{Department of Mechanical and Nuclear Engineering, Tennessee Tech University, Cookeville, TN 38505, USA}

\author{Danny Wang}
\affiliation{Department of Materials Science and Chemical Engineering, Stony Brook University, Stony Brook, NY, 11784, USA}

\author{David J. Sprouster}
\affiliation{Department of Materials Science and Chemical Engineering, Stony Brook University, Stony Brook, NY, 11784, USA}
\affiliation{Nuclear Reactor Laboratory, Massachusetts Institute of Technology, 77 Massachusetts Ave., Cambridge, MA 02139, USA}

\author{Shravan Venugopal}
\affiliation{Department of Mechanical and Nuclear Engineering, Tennessee Tech University, Cookeville, TN 38505, USA}

\author{Guiqiu Zheng}
\affiliation{Commonwealth Fusion Systems, Devens, MA 01434, USA}
\affiliation{Nuclear Reactor Laboratory, Massachusetts Institute of Technology, 77 Massachusetts Ave., Cambridge, MA 02139, USA}

\author{Dan Olds}
\affiliation{National Synchrotron Light Source II, Brookhaven National Laboratory, Upton, NY, 11973, USA}

\author{Jörg C. Neuefeind}
\affiliation{Neutron Scattering Science Directorate, Oak Ridge National Laboratory, Oak Ridge, Tennessee 37831-6475, United States}

\author{Stephen Lam}
\affiliation{Department of Chemical Engineering, University of Massachusetts Lowell, Lowell, Massachusetts 01854, United States}

\author{Boris Khaykovich}
\email{bkh@mit.edu}
\affiliation{Nuclear Reactor Laboratory, Massachusetts Institute of Technology, 77 Massachusetts Ave., Cambridge, MA 02139, USA}


\date{\today}%

\begin{abstract}
Several next-generation fission reactor designs employ molten salts such as FLiBe (\ce{2LiF-BeF2}), with some concepts using fuel dissolved directly in the salt. During operation, fission products such as cesium will accumulate in the salt mixture, potentially leading to an evolution of the thermophysical properties underpinned by the atomic structure. To understand the structural perturbations in FLiBe with 5 mol\% CsF, we conducted X-ray and neutron diffraction measurements, refined  empirical potential structure refinement (EPSR) models against the experimental data, and compared the resulting structure with neural network molecular dynamics (NNMD) simulations. Comparisons of the EPSR and NNMD structures distinguishes features constrained by the scattering data from those that remain model dependent. The new Cs-bearing correlations account for the changes in the total structure factor and pair-distribution function, while the FLiBe correlations remain minimally altered. We find that Cs slightly disrupts the intermediate-range fluoroberyllate network, increasing the fraction of free fluorine ions, while the local coordination remains largely unchanged. The Cs ions within FLiBe cluster extensively, with \ce{BeF4^{2-}} tetrahedra bridging neighboring cesium environments. In contrast to the minor structural perturbations in the liquid, the addition of  5 mol\% CsF suppressed the formation of the crystalline \ce{Li2BeF4} phase at room temperature, with the phase appearing only above \qty{180}{\degree C} upon heating. These experimentally constrained structural features provide a benchmark for atomistic models used to predict the behavior and properties of fission-product-containing FLiBe.
\end{abstract}

\maketitle


\section{Introduction}\label{sec:intro}
Molten salts are promising materials for use with advanced energy technologies such as nuclear reactors and concentrated solar power systems \citep{Roper2022_MoltenSaltAdvanced,Ito2005_AdvancesMoltenSalt,Williams2017_TechnologyAppliedRD}. Next-generation fission reactor designs are incorporating molten salts as coolants and fuel carriers \citep{Andreades2016_DesignSummaryMarkI,Serp2014_MoltenSaltReactor} whereas fusion reactors use salts in tritium breeding blankets \citep{Sorbom2015_ARCCompactHighfield,Forsberg2020_FusionBlanketsFluoridesaltcooled}. Salts are advantageous due to comparable heat transfer properties to water and low vapor pressures at elevated temperatures  \citep{Williams2008_EvaluationSaltCoolants}, enabling operation at much higher temperatures than water without the need for a pressure vessel. Molten-salt reactor concepts were first explored in the 1950s and 60s, when prototype molten salt reactors (MSRs) were designed, constructed, and operated to evaluate their feasibility \citep{Rosenthal1970_MoltenSaltReactorsHistory, Bettis1957_AircraftReactorExperiment,Robertson1965_MSREDesignOperations,MacPherson1985_MoltenSaltReactor}.  Among the most important challenges identified were controlling the salt chemistry and redox conditions to mitigate corrosion and materials degradation of salt-contacting alloys \citep{Sridharan2013_12CorrosionMolten,DeVan1962_CorrosionBehaviorReactor,Koger1972_EvaluationHastelloyAlloys,Koger1973_CorrosionProductDeposition}, and understanding fission-product behavior in the circulating salt, including their transport, deposition, and removal \citep{Compere1975_FissionProductBehavior, Riley2019_MoltenSaltReactor, Carlson2021_MoltenSaltReactors}. Thus, experimentally quantifying the fission-product-induced changes in the atomic structure of the circulating salt will be critical to accurately predicting the evolution of salt properties.

Halide salts have been the primary choice for MSRs, typically using either chlorides or fluorides. Optimal salt compositions vary depending on their intended application within reactors \citep{Serp2014_MoltenSaltReactor}.
While many compositions have merit, the most prominent for use as fuel-salt has become the 2:1 eutectic mixture of \ce{LiF-BeF2}, or FLiBe \citep{Williams2008_EvaluationSaltCoolants}, the focus of this study. In fuel-salt MSRs, fission products are produced during operations. As a result, online fuel reprocessing may be necessary since fission products could alter the physio-chemical properties of the salt \citep{Merk2018_DemandDrivenSalt, Riley2018_IdentificationPotentialWaste, Volkovich2003_TreatmentMoltenSalt}. These properties need to be well characterized to ensure operating conditions remain optimized and evolve as predicted. As such, the physical properties such as structure, density, viscosity, and others need to be simulated and experimentally verified. Since the atomic structure underpins thermophysical predictions, measurements of both the solid and liquid structure are essential to constrain and validate simulation. Of the fission products produced during operation, cesium is one of the most important due to relatively high fission yield \citep{Petruska1955_FISSIONYIELDSCESIUM}, having isotopes with long half lives, and being a potential health hazard for any release of \ce{^137Ce}. Cesium stability within molten salts has been shown to depend on its chemical form, including \ce{CsF}, \ce{CsCl}, and \ce{CsI}; the latter of which has been shown to be more volatile \citep{Benes2021_CesiumIodineRelease}.  

The liquid FLiBe structure is comprised of corner sharing \ce{BeF4^{2-}} tetrahedral units that assemble into short polymers, with \ce{Li} serving as a spacer between the segments. Bivalent alkali fluorides and chlorides are well-known network-forming liquids \citep{Rollet2011_StudiesLocalStructures}, with \ce{BeF2} notably forming a highly viscous and fully networked structure \citep{Cantor1969_ViscosityDensityMolten, Baes1970_PolymerModelBeF2, Smith2020_NewApproachCoupled}. This is in contrast to monovalent alkali fluorides, such as FLiNaK (eutectic mixture of \ce{LiF-NaF-KF}), which has only short-range ionic order without any persistent network formation \citep{Frandsen2020_StructureMoltenFLiNaK, Sprouster2022_MolecularStructurePhasea, Guo2023_XrayMolecularDynamics, Rakib2025_StructureMoltenF7LiNaK}.  The liquid structure of FLiBe was first measured using a lab X-ray diffractometer \citep{Vaslow1973_DiffractionPatternStructure} and then more recently our work using synchrotron X-rays and time-of-flight (TOF) neutron diffraction \citep{Fayfar2024_ComplexStructureMolten} presented the structure with the best resolution achievable. Molecular dynamics (MD) simulations have proved foundational to interpret the complex structure, using classical and ab initio MD (AIMD) \citep{Rahman1972_StructureMotionLiquid, Heaton2006_FirstPrinciplesDescriptionLiquid, Salanne2006_FirstprinciplesDescriptionLiquid, Nam2014_FirstprinciplesMolecularDynamics, Baral2021_TemperatureDependentPropertiesMolten, Winner2021_AbinitioSimulationStudies, Wang2022_FirstprinciplesMolecularDynamics}. Most recently, neural network molecular dynamics (NNMD) has achieved near AIMD accuracy at a fraction of the computational cost \citep{Lam2021_ModelingLiFFLiBe, Rodriguez2021_ThermodynamicTransportProperties, Li2024_CompositionalTransferabilityDeep, Zakiryanov2026_LiFBeF2MoltenMixtures}. 

The influence of dissolved impurity species such as corrosion and fission products have been investigated using both experimental and computational techniques. In FLiBe, the fission product europium was found to exhibit persistent clustering \citep{Li2026_AtomicScaleEuropiumSolute,Jiang2026_ConcentrationTransferableDeep}, and lanthanum was found to incorporate into and disrupt the fluoroberyllate network \citep{Moon2024_DensityMeasurementsMolten,Li2024_CompositionalTransferabilityDeep}. The actinide thorium forms strong eightfold fluorine complexes with minimal impact on the local FLiBe structure \citep{Wang2021_StructuresThoriumFluoride,Yin2025_LocalStructureIonic, Li2025_ChemicalFootprintsThorium}. For corrosion products, the chromium oxidation state has been shown to affect fluoride coordination \citep{Nam2014_FirstprinciplesMolecularDynamics}, while nickel exhibits a weak binding to chromium \citep{Attarian2024_StudiesNiCrComplexation}. In FLiNaK salt, cesium was found to form the room temperature crystalline phase \ce{CsLiF3} along with KF, LiF, and NaF phases, and in the liquid structure, the first structure factor peak at \qty{2}{\per\angstrom} increased in intensity \citep{Sprouster2022_MolecularStructurePhasea}. Molybdenum has a large variety of oxidation states, creating many complexes with fluorine \citep{Clark2021_ComplexationMoFLiNaK}. Corrosion experiments found that NiCr dissolution creates both metallic \ce{Ni^0} and \ce{Ni^{2+}} species \citep{Fayfar2023_InSituAnalysisCorrosion}. In chlorine salts, uranium \citep{Dai2018_MolecularDynamicsInvestigation} and plutonium \citep{Nguyen2023_ExploringNaClPuCl3Molten} were found to form large networks. Cesium mixed with \ce{UCl3-NaCl} salt was found to have a ten-fold coordination with chlorine, while pure \ce{CsCl} had a five- to six-fold coordination maximum \citep{Wang2025_CoordinationDrivenMixingBehavior}. Chromium formed octahedral chlorine complexes, forming intermediate-range networks \citep{Li2021_ComplexStructureMoltena}. Collectively, these studies demonstrate that the structural response to impurity species depends strongly on the identity, valence, and coordination chemistry of the dissolved species, with impurities forming diverse coordination complexes and participating in extended networks.

In this study, we measured the structure of FLiBe with 5 mol\% \ce{CsF} using synchrotron X-ray and neutron diffraction. To uncover the structure from experimental measurements, we performed empirical potential structural refinement (EPSR) methods, enabling experimentally constrained analysis of the coordination numbers, cluster sizes, and bond angle distributions \citep{Soper1996_EmpiricalPotentialMonte,Soper2001_TestsEmpiricalPotential,Soper2005_PartialStructureFactors}. Furthermore, we performed NNMD simulations. This combination allows for a deep understanding of the liquid structure of \ce{FLiBe-CsF}. We found that the addition of Cs does not significantly disrupt the high-temperature liquid structure of FLiBe, which characteristically contains a \ce{BeF4^{2-}} tetrahedral network; instead, \ce{Cs-F} polyhedra tend to cluster. Since monovalent cesium has a weakly bound coordination environment, we used the relative angular distance (RAD) method to determine the coordination shell \citep{Higham2016_LocallyAdaptiveMethod,Higham2018_OvercomingLimitationsCutoffs}, separating out atoms belonging to secondary shells. By thoroughly investigating the atomic structure, we have benchmarked molecular-dynamic simulations of liquid \ce{FLiBe-CsF} against experimentally constrained EPSR models, establishing which structural features can be resolved by scattering measurements. In contrast to the minimal changes in the liquid structure, the room-temperature crystal structure of FLiBe (\ce{Li2BeF4}) is completely destroyed by the addition of 5 mol\% CsF.

\section{Experimental and computational methods}\label{sec:methods}

\subsection{Sample preparations}\label{sec:methods:sample}
A 2:1 stoichiometric ratio of LiF and \ce{BeF2} was prepared from constituent materials from Materion salt. The FLiBe salt was mixed with 5 mol\% CsF. For X-ray diffraction measurements, the salt samples were loaded into sealed graphite capsules within flame sealed NMR quartz tubes as described in refs. \citep{Sprouster2022_MolecularStructurePhasea,Fayfar2024_ComplexStructureMolten}. The salt was premelted into the graphite capsule to achieve a maximal packing fraction of salt in the beam. The neutron samples were prepared using isotopically enriched \ce{^7LiF} provided by Oak Ridge National Laboratory. The salt was loaded into \qty{5}{mm} vanadium canisters and sealed with a flanged titanium lid with a graphite-foil gasket. 

\subsection{X-ray and Neutron scattering}\label{sec:methods:scattering}
X-ray diffraction (XRD) experiments were performed at the PDF beamline (28-ID-1) at the National Synchrotron Light Source II (NSLS-II) \citep{Ivashkevych2020_HardXRayPair}. The samples were encapsulated in capped graphite containers within sealed \qty{5}{mm} quartz tubes as described in refs. \citep{Sprouster2022_MolecularStructurePhasea, Fayfar2024_ComplexStructureMolten}. The samples were heated using a hot-air blower, starting from room temperature to beyond the melting point. The graphite peaks from the sample holder were utilized to calibrate the sample temperature. The beamline uses a Perkin Elmer amorphous-silicon-based flat-panel detector system, which translates between a short-distance high-$Q$ position and a far-distance high-resolution position. The samples-to-detector distances and detector orientations were calibrated by measuring a \ce{LaB6} standard sample. The raw images were dark current corrected and the beamstop and artifacts were masked prior to performing azimuthal integrations using pyFAI \citep{Kieffer2013_PyFAIVersatileLibrary}. The diffraction patters of the samples were background corrected by subtracting patterns obtained of the empty sample container. The residuals of the graphite peak subtraction were removed and smoothed. LiquidDiffract was used to reduce the scattering intensity to the structure factor $S(Q)$ and calculate the pair distribution function (PDF) \citep{Heinen2022_LiquidDiffractSoftwareLiquid}. The wave-vector transfer $Q$ for elastic scattering is given by $Q=\frac{2 \pi}{d}=\frac{4 \pi}{\lambda}\sin{\theta}$, where $d$ is the interatomic distance, $\lambda$ is the wavelength, $\theta$ is half the scattering angle $2\theta$. The X-ray FLiBe measurements were presented in \textcite{Fayfar2024_ComplexStructureMolten}. However, the raw measurements were reprocessed and reduced using LiquidDiffract, which more reliably reduces the scattering measurements \citep{Gallington2023_ReviewCurrentSoftware}. Matching data reduction procedures were used for both the \ce{FLiBe-CsF} and FLiBe X-ray diffraction measurements. 

Neutron diffraction (ND) measurements were performed at the NOMAD beamline at the Spallation Neutron Source (SNS) \citep{Neuefeind2006_NanoscaleOrderedMaterials}. The samples were sealed in standard \qty{5}{mm} vanadium cans and installed in the ILL vanadium vacuum furnace. The samples were measured continuously from room temperature up to \qty{700}{\degree C}. During heating, the sample temperatures were  held at \qtylist{550;700}{\degree C} for a 7 hour acquisition each. The raw scattering measurements were normalized using the vanadium standard and background subtracted using measurements of the empty sample container. The ADDIE software suite was used to calculate the structure factor and PDF from the scattering intensity \citep{McDonnell2017_ADDIEADvancedDIffraction}. 

The static structure factor $S(Q)$ is composed of partial structure factors that depend on the atomic composition of the material and the scattering probe (X-rays or neutrons)
\begin{align}\label{eq:SofQ_partials}
    \begin{split}
        S^{(\textrm{n}/\textrm{x})}(Q) 
        &= \sum_{\alpha,\beta \geq \alpha} S^{(\textrm{n}/\textrm{x})}_{\alpha \beta} (Q) \\
        &= \sum_{\alpha,\beta \geq \alpha} w^{(\textrm{n}/\textrm{x})}_{\alpha \beta}(Q) S_{\alpha \beta}(Q),
    \end{split}
\end{align}
with 
\begin{equation}\label{eq:weights_neutron_xray}
    w^{(\textrm{n}/\textrm{x})}_{\alpha \beta}(Q) = \frac{c_\alpha c_\beta f_\alpha(Q) f_\beta(Q)}
    {\left[ \sum_\alpha c_\alpha f_\alpha(Q) \right]^2} \left( 2 - \delta_{\alpha\beta} \right),
\end{equation}
where $\alpha$ and $\beta$ denote the salt species, $c$ is the corresponding concentration, $f(Q)$ is the species-dependent scattering factor, taken as the X-ray atomic form factor for X-ray scattering and as the $Q$-independent neutron scattering length for neutron scattering, and $\delta_{\alpha\beta}$ is the Kronecker delta. Performing a Fourier-sine transform on the structure factor yields the PDF
\begin{equation}\label{eq:gofr_fourier}
    g(r) - 1 = \frac{1}{2 \pi^2 \rho_0} \int_Q^{Q_\text{max}} Q \left[ S(Q) - 1 \right] \frac{\sin{Qr}}{r} d Q,
\end{equation}
where $\rho_0$ is the atomic number density. The PDF is the ratio of the atomic density a distance $r$ around a central atom relative to the bulk density \citep{Neuefeind2002_HighEnergyXRD,Fischer2006_NeutronXrayDiffraction,Peterson2021_IllustratedFormalismsTotal,Zhou2022_MolecularDynamicsSimulation}. 

The coordination number (CN) in studies of disordered systems is commonly obtained by integrating the partial PDF $g_{\alpha\beta}(r)$ to the first minimum following the first pair-correlation peak. This $g(r)$ cutoff (GC) method is given by
\begin{equation}\label{eq:CNs}
    N(r_{\textrm{min}}) 
    = 4 \pi \rho_\beta \int_0^{r_{\textrm{min}}} r^2 g_{\alpha\beta}(r) dr,
\end{equation}
where $\rho_\beta=c_\beta\rho_0$ is the number density of the $\beta$ atoms surrounding the central $\alpha$ atom, and $r_{\textrm{min}}$ is the first minimum in the partial PDFs. This approach provides a convenient measure of the number of atoms of type $\beta$ within a radial shell around a central atom of type $\alpha$. However, in multicomponent ionic liquids, this radial-shell population does not necessarily correspond to direct nearest-neighbor coordination. This distinction is especially important for like-charge pairs, where the first pairwise correlation peak may reflect second-shell ordering. Therefore, coordination numbers were also calculated using the relative angular distance (RAD) method, which defines neighbors from the instantaneous local geometry by excluding atoms that are blocked by closer intervening atoms. In the RAD method, a candidate neighbor $j$ is included in the coordination shell of atom $i$ only if it is not blocked by any closer atom $k$, which is evaluated using the criterion
\begin{equation}
    \frac{1}{r_{ij}^{2}} >
    \frac{1}{r_{ik}^{2}}\cos\theta_{jik}.
    \label{eq:rad}
\end{equation}
where $r_{ij}$ and $r_{ik}$ are distances from the central atom $i$, and $\theta_{jik}$ is the angle subtended at $i$ by atoms $j$ and $k$ \citep{Higham2016_LocallyAdaptiveMethod,Higham2018_OvercomingLimitationsCutoffs}. To compare with previous literature values using the GC method and directly compare the two methods, both are reported.  

\subsection{Molecular-dynamics simulations}\label{sec:methods:simulations}
\subsubsection{Ab-Initio Molecular Dynamics Simulations (AIMD)}
The AIMD simulations for FLiBe-4.9\% CsF were performed under Born-Oppenheimer approximation using Vienna Ab-Initio Simulation Package (VASP) \citep{Kresse1996_EfficientIterativeSchemes}. In VASP, generalized-gradient-approximation (GGA) in the form of Perdew-Burke-Ernzerhof (PBE) exchange correlation functional \citep{Perdew1996_GeneralizedGradientApproximation} was adopted for electronic self-consistent calculations. Projector augmented wave (PAW) pseudopotentials recommended by VASP were used to model core electrons for Be (\ce{2s^2}), Cs\_sv (\ce{5s^2 5p^6 6s^1}), Li\_sv (\ce{1s^2 2s^1}), and F (\ce{2s^2 2p^5}). A large plane wave cutoff of 600 eV with a \qty{1e-5}{eV} convergence criterion for electronic self-consistent steps. Calculations were performed using gamma point. The parameters chosen yield convergence within 2 meV per atom. The density functional theory (DFT)-D3 formulation proposed by Grimme \citep{Grimme2010_ConsistentAccurateInitio} was used to account for the dispersion interactions. The canonical ensemble (NVT) using a Nosé-Hoover thermostat \citep{Hoover1985_CanonicalDynamicsEquilibrium} was employed while maintaining the periodic boundary conditions. In all AIMD simulations, a time step of 2 femtosecond (fs) was used.
To simulate FLiBe-4.9\% CsF in AIMD, a supercell containing 95 atoms (13 Be, 26 Li, 54 F, and 2 Cs) was used. At both \qtylist{510;700}{\degree C}, the simulation cell volume for FLiBe-CsF system was estimated by adding the equilibrium volume of the pure FLiBe and the CsF as taken from their experimental temperature-density relation \citep{Janz_MoltenSaltsHandbook,Zaghloul2003_ThermoPhysicalPropertiesEquilibrium}. This yielded the estimated density of \qty{2.24}{g/cc} at \qty{510}{\degree C}, while density was estimated to be \qty{2.13}{g/cc} at \qty{700}{\degree C}. At the start of the AIMD calculations, the supercell was first initialized with a random structure generated with the Packmol package to avoid overlap in atomic positions \citep{Martinez2009_PACKMOLPackageBuilding}. This structure was further equilibrated for ~94 picosecond (ps) and ~67 ps at \qtylist{510;700}{\degree C}, respectively to allow sufficient time to develop and equilibrate the liquid structure. According to our previous work, including FLiBe \citep{Fayfar2024_ComplexStructureMolten} and other multicomponent salts \citep{Chahal2022_TransferableDeepLearning}, a simulation trajectory of ~50 ps should be sufficient to sample diverse molten salt configurations to train a reliable NNIP model.

\subsubsection{NNIP Training}
The DeePMD-kit (DP-kit) package (version 1.3.3) was employed \citep{Wang2018_DeePMDkitDeepLearning} to train a smooth and continuously differentiable potential energy surface using Deep-Pot-Smooth Edition (DeepPot-SE) model \citep{Zhang2018_EndtoendSymmetryPreserving}. The DeepPot-SE model learns a mapping between local environment of each atom within a hard cut-off radius of 8 Å to a per-atom energy, such that the sum of atomic energies for each configuration corresponds to its reference DFT energy. Thereafter, the atomic forces are computed using gradients of NNIP-predicted energies. Both the reference energies and forces are included to evaluate the loss function which is minimized during training of an DeepPot-SE model. Our training dataset comprised 11,019 configurations at \qty{510}{\degree C} and 10,283 configurations at \qty{700}{\degree C} from the AIMD simulations of FLiBe-4.9\% CsF system containing 95 atoms each. During the training, the AIMD datasets were shuffled and were split in 80\% and 20\% for training and validation, respectively. A smooth cutoff radius of 2 Å was chosen, and embedding network and fitting network size of \{25,50,100\} and \{240,240,240\}, respectively were chosen. The tunable prefactors in loss function were chosen as 0.002, 1000, 1, 1 for p\_\{e,start\}, p\_\{f,start\}, p\_\{e,limit\}, and p\_\{f,limit\}, respectively. These hyperparameters previously resulted in a well-fitted potential energy surface for multicomponent molten salt system, including FLiBe \citep{Fayfar2024_ComplexStructureMolten,Chahal2022_TransferableDeepLearning}. The NNIP training was continued up to 500,000 steps using a batch size of 5. The trained NNIP model yielded energy and force errors of 3.85 meV/atom and 22.79 meV/Å, respectively on the validation set. As such, the low energy and force validation errors suggest a well-fitted potential energy surface for performing NNIP-based MD simulations of FLiBe-4.9\%CsF system in the next section. 

\subsubsection{Neural Network Molecular Dynamics }
The trained NNIP for \ce{FLiBe-CsF} was used in Large Scale Atomic/Molecular Massively Parallel Simulator (LAMMPS) via the interface with DeePMD-kit \citep{Plimpton1995_FastParallelAlgorithms}. Larger systems were simulated in NNIP-based molecular dynamics (NNMD) simulations in order to sufficiently represent the intermediate-range structure. Specifically, periodic systems containing 2565 atoms were simulated at \qtylist{510;700}{\degree C} using cell parameters of 31.7 Å and 32.36 Å, respectively. Each simulation cell contained 702 Li, 351 Be, 54 Cs, and 1458 F atoms at densities corresponding to the estimated experimental densities obtained from the procedure described in the previous section. During NNMD simulations, at least 12.5 nanosecond (ns) simulations were performed using 2 fs time step employing Nosé-Hoover thermostat \citep{Hoover1985_CanonicalDynamicsEquilibrium} in constant volume ensemble (NVT). Periodic boundary conditions were maintained in all three directions. The equilibrated trajectory longer than 6.5 ns was used at both temperatures to evaluate pair distribution functions, coordination states, structure factor, as well as to perform their cluster analysis. More details and discussion on each analysis are provided in the following sections.

\subsection{Empirical potential structural refinement}
Reverse Monte Carlo (RMC) techniques are powerful for uncovering the atomic structure of disordered systems from experimental diffraction measurements. For this study, we employed the empirical potential structural refinement (EPSR) \citep{Soper1996_EmpiricalPotentialMonte,Soper2001_TestsEmpiricalPotential,Soper2005_PartialStructureFactors} technique, which bridges conventional atomistic simulations and RMC approaches by incorporating both interatomic potentials and experimental diffraction constraints. It starts by building a large randomized box of atoms (exact atoms counts listed in Table S1) that interact through an initial reference potential, comprised of a Coulomb potential, an exponential repulsion, and the 12-6 Leonnard-Jones potential. The structure factor and the PDF are calculated from the atomic configuration and compared with experimental measurements. The configuration is equilibrated using a Metropolis Monte Carlo method, which involves randomly moving each atom and evaluating the energy change \citep{Metropolis1953_EquationStateCalculations}. Iteratively, the reference potential parameters are tuned and the system equilibrates until the local structure has a reasonable agreement with the experimental measurements. Then the empirical potential is enabled and slowly increased in amplitude to fit the atomic configuration to the experimental measurements. 

Once the EPSR method has converged with the experimental data, the resulting atomic configurations are ensemble-averaged, thereby enabling the calculation of structural descriptors such as partial PDFs, bond-angle distributions, coordination numbers, among other related quantities. While this technique is very powerful in uncovering the partial PDFs without the need for expensive experimental techniques such as isotopic substitution using neutron diffraction, it does not necessarily uncover a unique structural solution. However, this can be improved by constraining the refinement with additional sources of experimental measurements (XRD and ND). Since we conducted both XRD and ND measurements on \ce{FLiBe-CsF} and FLiBe, we used both datasets as inputs during the EPSR refinement. The parameters from our EPSR refinements are shown in Table S1 and S2. 

\begin{figure*}[tb]
    \centering
    \includegraphics[]{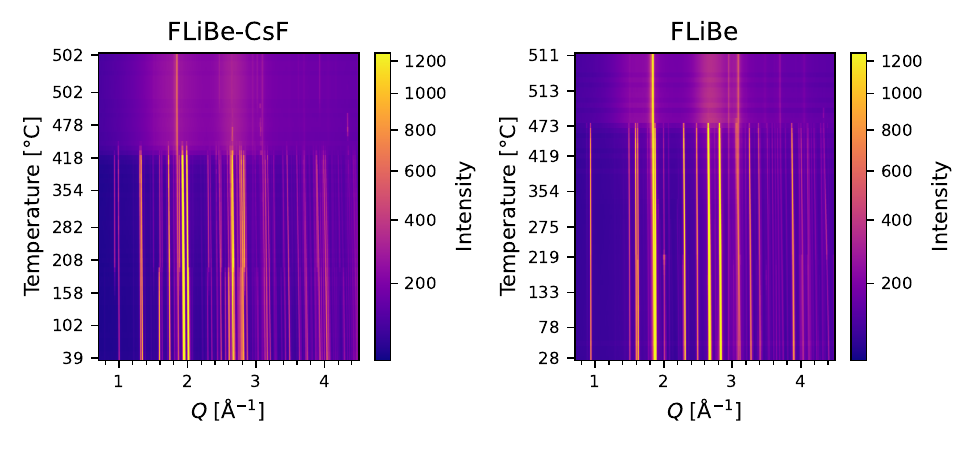}
    \caption{X-ray diffraction heatmaps of \ce{FLiBe-CsF} and \ce{FLiBe} showing the pattern change with temperature. The intensity is shown on a square root scale. After melting, the only peaks remaining are from the graphite sample holder. Note, the y-scales are not linear, but rather interpolations of the experimental heating profiles.}
    \label{fig:xrd_heating}
\end{figure*}

\subsection{Quantifying agreement between models and experiments}
To quantify the agreement between the EPSR and NNMD models with experimental data, we calculated $R_\chi$  \citep{Wright1994_NeutronScatteringVitreous,Du2009_MolecularDynamicsSimulation,Alderman2015_LiquidB2O31700,Wilson2016_StructureDynamicsNetworkforming,Zhou2021_ExperimentalMethodQuantify}, defined by
\begin{equation}\label{eq:R_factor}
    R_\chi = \left( \sum_i \left[f_{\textrm{exp}}(x_i) - f_{\textrm{sim}}(x_i) \right]^2 / \sum_i f_{\textrm{exp}}^2(x_i) \right)^{1/2},
\end{equation}
where  $f_{\textrm{exp}}$ and $f_{\textrm{sim}}$ correspond to measured and simulated $S(Q)$ and $g(r)$. The structure factor was evaluate from $0.5 \leq Q \leq$\qty{15}{\per \angstrom}, and the PDF from $0.5 \leq r \leq$\qty{10}{\angstrom}. The resulting calculations are shown in Fig. S16.

\section{Experimental and computational results}\label{sec:results}
\subsection{Solid Structure}\label{sec:results:solid}
\begin{figure}[bt]
    \centering
    \includegraphics[]{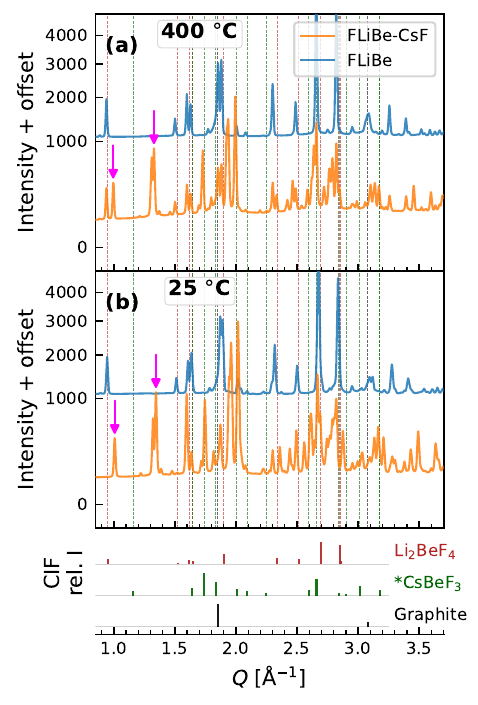}
    \caption{X-ray diffraction patterns of \ce{FLiBe-CsF} and \ce{FLiBe} (a) at \qty{400}{\degree C} and (b) near room temperature. Predicted XRD reflections for \ce{Li2BeF4}, \ce{CsBeF3}, and graphite are included at the bottom of the figure with corresponding dashed vertical lines. The \ce{CsBeF3} reflections did not match with either experimental diffraction pattern. Magenta arrows highlight the strongest Cs-containing reflections that do not correspond to predicted phases.}
    \label{fig:xrd_solid_rt_hightemp}
\end{figure}

The evolution of the XRD pattern during heating for \ce{FLiBe-CsF} and FLiBe is presented in \cref{fig:xrd_heating}. The FLiBe heating measurements show melting completing around \qty{470}{\degree C} \textcite{Fayfar2024_ComplexStructureMolten}, while \ce{FLiBe-CsF} melts at a lower temperature at around \qty{450}{\degree C}. XRD patterns at room temperature and \qty{400}{\degree C} are shown in \cref{fig:xrd_solid_rt_hightemp}. At room temperature, the \ce{FLiBe-CsF} mixture exhibits numerous sharp reflections arising from one or more crystalline Cs-containing phases. These reflections could not be assigned uniquely to a single crystalline phase. This is in part due to the structural complexity and limited reference data available for \ce{Cs-Li-Be-F} compounds, and highly spotty diffraction rings, indicative of coarse crystallites formed during cooling. The course crystalline likely resulted from the melt-solidification step during sample preparation. The specimens were initially loaded into capsules as powders and subsequently melted to increase the packing fraction. They resolidified as polycrystalline specimens, which are not ideal for phase identification using XRD. In addition, the samples were allowed to cool naturally, potentially preventing phases enough time to crystallize. Because the objective of the present work was to examine and follow the structural evolution during heating and melting, rather than to determine the crystal chemistry of the Cs-bearing reaction products, identifying these reflections are not discussed further. Three of the strongest Cs-containing reflections are highlighted by the magenta arrows in \cref{fig:xrd_solid_rt_hightemp} for reference. 

Upon heating above approximately \qty{180}{\degree C}, additional reflections corresponding to crystalline FLiBe emerge while the unidentified Cs-containing reflections remain present. This indicates that FLiBe crystallizes or becomes sufficiently ordered while the Cs-bearing phases persist over the same temperature interval. Both sets of reflections disappear upon melting at approximately \qty{450}{\degree C}. Consequently, the principal conclusions regarding the melting behavior of the FLiBe–CsF mixture are independent of the precise assignment of these secondary reflections. The emergence of the FLiBe reflections during heating suggests that a fraction of the FLiBe may initially exist in a poorly ordered or highly strained state, or that partial solid-state re-equilibration occurs upon heating. Distinguishing between these possibilities would require complementary structural characterization and is beyond the scope of the present study.

\subsection{Liquid Structure}\label{sec:resuls:liquid}
\subsubsection{Interpreting total structure factor and PDF}
\begin{figure*}[bt]
    \centering
    \includegraphics[]{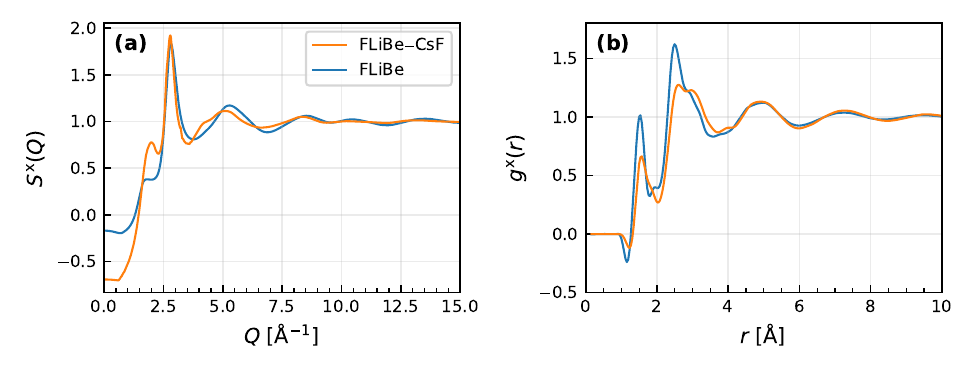}
    \caption{(a) The X-ray total structure factor and (b) total pair distribution function for \ce{FLiBe-CsF} (blue) and \ce{FLiBe} (orange).}
    \label{fig:xray_csflibe_flibe_liquid}
\end{figure*}

The liquid X-ray structure factors and PDFs of \ce{FLiBe-CsF} and FLiBe are presented in \cref{fig:xray_csflibe_flibe_liquid}. The \ce{FLiBe-CsF} and FLiBe measurements were performed at \qtylist{500;510}{\degree C}, respectively. With the inclusion of Cs, the first peak in the structure factor at \qty{1.8}{\per\angstrom} increases in intensity and shifts to higher $Q$. The PDF has a reduction in the first peak at \qty{1.55}{\angstrom} and the third peak at \qty{2.5}{\angstrom}. 
The neutron structure factor and PDF are included in Figure S1. X-ray diffraction provides significantly more contrast of the Cs-induced structural variations in FLiBe than neutron diffraction.

\begin{figure*}[tb]
    \centering
    \includegraphics[]{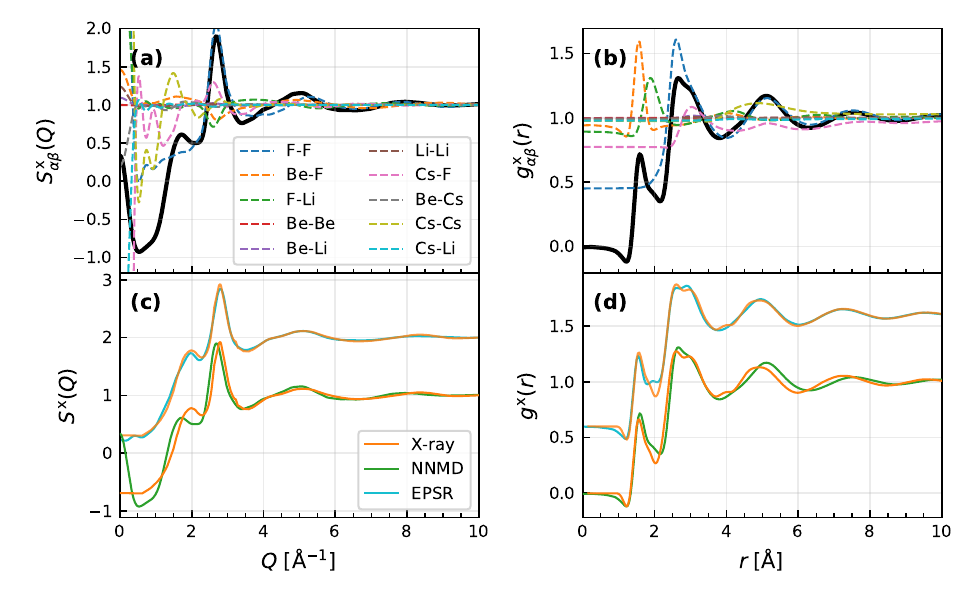}
    \caption{The experimental and simulated structure factor and pair distribution function of \ce{FLiBe-CsF} at \qty{500}{\degree C}. (a) The X-ray weighted partial structure factor and (b) partial pair distribution function calculated from NNMD simulations. (c) The X-ray total structure factor and (d) total pair distribution function from experiments, NNMD simulations, and EPSR models.}
    \label{fig:csflibe_xray_nnmd}
    \phantomsubcaption\label{fig:csflibe_xray_nnmd_a}
    \phantomsubcaption\label{fig:csflibe_xray_nnmd_b}
    \phantomsubcaption\label{fig:csflibe_xray_nnmd_c}
    \phantomsubcaption\label{fig:csflibe_xray_nnmd_d}
\end{figure*}

The predicted NNMD partial and total structure factors and PDFs of \ce{FLiBe-CsF} are displayed in \cref{fig:csflibe_xray_nnmd}, revealing the pair contributions to the total structure and PDF peaks. The total structure factor and PDF from the EPSR model fit to the experimental data is included in \cref{fig:csflibe_xray_nnmd_c,fig:csflibe_xray_nnmd_d} and the partials are presented in Fig. S2. When adding Cs to FLiBe, the change in the first structure factor peak occurs due to the \ce{Cs-F} and \ce{Cs-Cs} correlations combining with the \ce{Be-F} and \ce{F-F} partial peaks. The change in the PDF around \qty{3}{\angstrom} is attributed to the inclusion of the \ce{Cs-F} pair distribution. The \ce{Li-F} peak remains largely unchanged with a very slight increase in intensity. To visualize the structural changes Cs had in each of the partials, Figures S9 and S10 presents the unweighted partial PDFs and structure factors of both compositions from EPSR and NNMD. The FLiBe pair partials comprised of \ce{Li-F-Be} change very little with the inclusion of 5 mol\% CsF. As such, the change in the total structure factor is dominated by the new Cs correlations rather than perturbing the FLiBe structure substantially. 

\subsubsection{Local structure and CNs}
The calculated first peak distances and average coordination number values are arranged in \cref{tab:coords_and_peak}. Refining EPSR models against the experimental measurements enables straightforward calculations of the distances and CNs using the accumulated atomic configurations. From the FLiBe EPSR model, the CN$_\textrm{GC}$ for \ce{Be-F} is $4.021(20)$, much closer to the predicted value of 4 than the 3.73(15) value determined through direct Gaussian fitting in \textcite{Fayfar2024_ComplexStructureMolten} using the same measurements. Calculating the experimental coordination number through Gaussian fitting works best when the peaks are isolated, such as in \ce{BeF2} and \ce{SiO2} \citep{Wright1989_NeutronDiffractionMolecular}. Since FLiBe has overlapping \ce{Be-F} and \ce{Li-F} peaks, the Gaussian fitting method has greater uncertainties, which resulted in the coordination number being below this current value obtained from EPSR models. 

\begin{figure*}[tb]
    \includegraphics[]{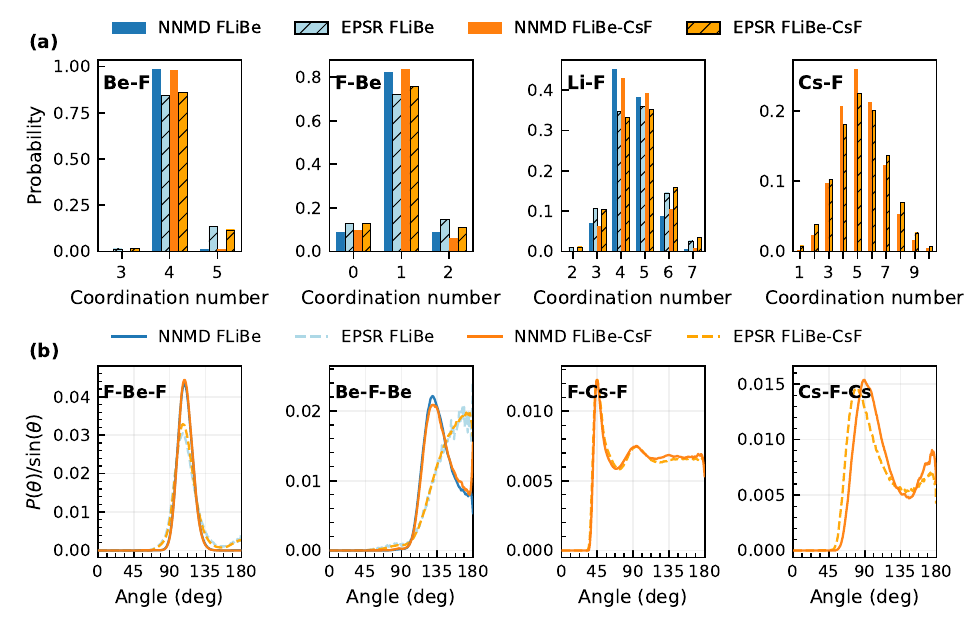}
    \caption{(a) CN distributions calculated using the RAD method, and (b) angle distribution functions normalized by $\sin \theta$.}
    \label{fig:CN_dist_bond_angles}
    \phantomsubcaption\label{fig:CN_dist_bond_angles_a}
    \phantomsubcaption\label{fig:CN_dist_bond_angles_b}
\end{figure*}

The CNs are presented using the RAD method in addition to the commonly used GC method. The RAD method determines the CNs without any arbitrary parameters such as the $r_\textrm{min}$ value used in the GC calculation. Instead, it uses a geometric relation with the accumulated atomic configurations directly. Nearest atoms are located around a central atom and are excluded from the first coordination shell if blocked by closer atom using the criterion in \cref{eq:rad}. For opposite-charged pairs such as \ce{Be-F}, the methods  using NNMD simulations of FLiBe produce similar results: CN$_\textrm{GC}=3.9950(19)$ and CN$_\textrm{RAD}=4.0115(8)$. From EPSR, the RAD method calculated a slightly higher CN$_\textrm{RAD}=4.134(4)$ compared with CN$_\textrm{GC}=4.021(20)$. This is likely caused by EPSR producing a less well-defined \ce{[BeF4]^{2-}} tetrahedron, evidenced by a broader \ce{F-Be-F} angle distribution function (ADF) shown in \cref{fig:CN_dist_bond_angles_b}. For like-charged pairs such as \ce{Be-Be}, the GC method differs greatly from the RAD method, as has been discussed previously \citep{Higham2018_OvercomingLimitationsCutoffs}. Within FLiBe, Be atoms do not directly coordinate to each other, but rather, they are linked together by a bridging F. The RAD method accurately captures this, resulting in a CN$_\textrm{RAD}$ near zero. While the GC method relies on the $r_\textrm{min}$ value, which for \ce{Be-Be} and other like-charge pairs, is beyond the first coordination shell.

\begin{table*}[tb]
    \centering
    \caption{First-peak distances and coordination numbers for FLiBe and \ce{FLiBe-CsF} determined from NNMD simulations and EPSR models fit to X-ray and neutron diffraction measurements. First-peak distances were defined as the maxima of skewed-Gaussian fits to the partial \(g_{\alpha\beta}(r)\) functions. Coordination numbers are obtained using the GC method with the listed $r_{\min}$ cutoff and the RAD method using \cref{eq:rad}. First-peak-distance uncertainties combine the standard error across trajectory blocks with the fit-parameter covariance in quadrature. Average CN$_\mathrm{GC}$ uncertainties combine the block standard error with sensitivity to the fixed integration cutoff. RAD CN uncertainties are the standard deviation of the blockwise mean coordination numbers. }

    \begin{ruledtabular}
    \begin{tabular}{lllllllll}
        Composition & Pair 
        & \multicolumn{2}{c}{First peak distance (\unit{\angstrom})}
        & $r_{\min}$ (\unit{\angstrom})
        & \multicolumn{2}{c}{CN$_{\mathrm{GC}}$}
        & \multicolumn{2}{c}{CN$_{\mathrm{RAD}}$} \\ 
        \cmidrule(lr){3-4} \cmidrule(lr){6-7} \cmidrule(lr){8-9}
        
        & & NNMD & EPSR & & NNMD & EPSR & NNMD & EPSR \\ 
        \midrule
        
        \multirow{4}{*}{FLiBe}
        & \ce{Be-F}  & 1.5500(5)  & 1.5127(4)   & 2.35 & 3.9950(19) & 4.021(20) & 4.0115(8)   & 4.134(4) \\
        & \ce{Li-F}  & 1.8650(25) & 1.8781(9)   & 2.75 & 4.68(12)   & 4.82(17)  & 4.504(5)    & 4.613(7) \\
        & \ce{F-F}   & 2.5792(6)  & 2.5163(5)   & 4.00 & 12.3(3)    & 12.0(4)   & 0.060(3)    & 0.2113(30) \\
        & \ce{Be-Be} & 2.946(18)  & 3.010(22)   & 3.45 & 0.705(16)  & 1.17(5)   & 0.000003(9) & 0.0039(16) \\
        
        \midrule
        
        \multirow{8}{*}{FLiBe--CsF}
        & \ce{Be-F}  & 1.55002(15) & 1.5125(13) & 2.35 & 3.9926(13) & 4.016(16) & 4.0084(5)   & 4.111(7) \\
        & \ce{Li-F}  & 1.863(3)    & 1.8707(7)  & 2.75 & 4.78(12)   & 4.90(20)  & 4.5643(25)  & 4.664(13) \\
        & \ce{F-F}   & 2.5847(21)  & 2.5359(5)  & 4.00 & 11.9(3)    & 11.6(3)   & 0.0663(9)   & 0.2010(21) \\
        & \ce{Be-Be} & 2.957(13)   & 3.013(15)  & 3.45 & 0.528(15)  & 0.87(4)   & 0.000030(8) & 0.0023(6) \\
        & \ce{Cs-F}  & 3.0323(19)  & 3.040(8)   & 4.20 & 9.28(28)   & 10.0(3)   & 5.232(14)   & 5.30(6) \\
        & \ce{Cs-Be} & 3.8510(23)  & 3.908(18)  & 5.35 & 5.79(8)    & 6.44(13)  & 0.0225(8)   & 0.059(5) \\
        & \ce{Cs-Li} & 4.167(5)    & 4.12(6)    & 5.40 & 4.54(14)   & 7.84(27)  & 0.0601(17)  & 0.496(26) \\
        & \ce{Cs-Cs} & 4.932(19)   & 4.32(8)    & 6.65 & 7.05(13)   & 3.85(13)  & 0.0144(8)   & 0.069(14) \\
        
    \end{tabular}
    \end{ruledtabular}
    
    \label{tab:coords_and_peak}
\end{table*}

The FLiBe structure itself remains largely unchanged with the addition of 5 mol\% CsF as seen by the first peak distances and CN values. The \ce{Be-F}, \ce{Li-F}, and \ce{F-F} first peak distances and CNs vary only slightly with the addition of Cs. The measurable changes in the total X-ray PDF come from the addition of the \ce{Cs-F} partial PDF, which has a first peak distance of \qty{3.040(8)}{\angstrom} and a CN$_\textrm{RAD}=5.30(6)$ from EPSR fits. The GC method determines the value to be $10.0(3)$ using a $r_\textrm{min}$ value of 4.20. The RAD method for \ce{Cs-F} excludes many neighboring atoms that have intermediary blocking atoms, which are not accounted for in the GC method. Furthermore, the GC method only considers the two coordinated species, Cs and F in this case, and many of the blocking atoms are Be or Li. Therefore, the first peak in the \ce{Cs-F} partial PDF is comprised of the first and second coordination shells, according to the RAD method. 

The individual CN distributions shown in \cref{fig:CN_dist_bond_angles_a} provide more detail of the structural variations caused by CsF. The \ce{Be-F} structure is still dominated by fourfold-coordinated tetrahedra with a few more fivefold-coordinated species found from EPSR. The \ce{F-Be} coordination, the average number of Be atoms surrounding a central F, is a marker for the amount of \ce{BeF4^{2-}} network participation mediated by bridging fluorine.  Adding CsF causes a decrease in the fluorine atoms coordinated to two Be atoms and increases single coordinations. The \ce{BeF4^{2-}} tetrahedra form networks predominantly through corner sharing. Although, a small amount of edge sharing is present as well, as shown directly in Fig. S15 and indirectly by the small peak in the \ce{Be-F-Be} ADF at \qty{90}{\degree} in \cref{fig:CN_dist_bond_angles_b}, matching other studies \citep{Zakiryanov2026_LiFBeF2MoltenMixtures}. The zero coordinated \ce{F-Be} species are the fluorine atoms not participating in the fluoroberyllate network, which are only weakly bound to lithium atoms. In this case, the \ce{F-Li} species have such low cage-correlation times that the fluorine atoms not participating in the \ce{BeF4^{2-}} network are considered free fluorine atoms, a proxy for the fluoracidity \citep{Winner2021_AbinitioSimulationStudies,Hunsberger2026_AcidbaseEquilibriaMolten}. NNMD simulations show a 14.5\% increase in the amount of free fluorine atoms with the inclusion of Cs, while EPSR does not predict any significant change. For both NNMD and EPSR, the \ce{Li-F} CN distribution changes very little with the addition of Cs. In both calculations, the \ce{Cs-F} coordination shows a fairly broad distribution with a maximum at fivefold coordinated species, extending to ninefold-coordinated species. The \ce{F-Cs-F} ADF in \cref{fig:CN_dist_bond_angles_b} is fairly broad, with maxima at \qty{45}{\degree} and \qty{92}{\degree} that do not correspond to any well-defined polyhedron geometry. We note that nearly all F ions within the first coordination shell of Cs also participate in \ce{BeF4^{2-}} tetrahedra.

\begin{figure*}[tb]
    \centering
    \includegraphics[]{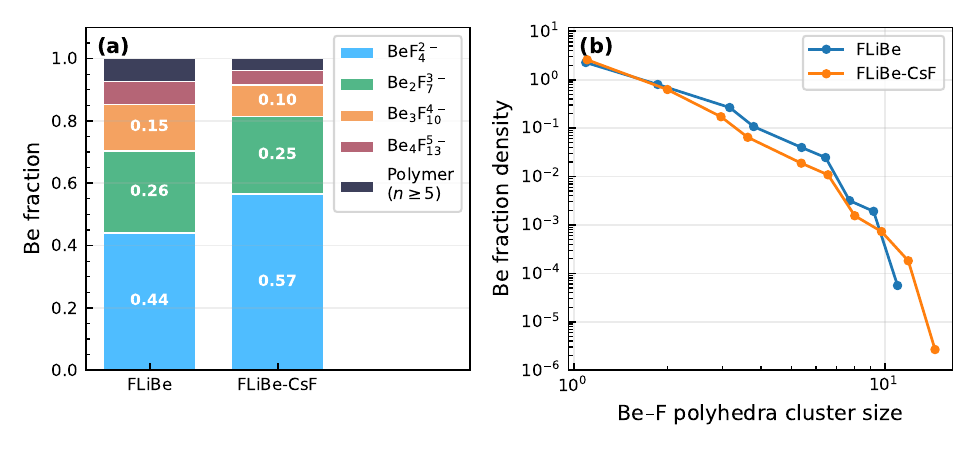}
    \caption{(a) The fraction of Be atoms participating in monomeric, oligomeric, and polymeric \ce{Be-F} tetrahedral species, and (b) the Be-weighted cluster-size distribution from NNMD simulations at \qty{510}{\degree C}. Both figures show the fraction of Be atoms within clusters of size $n$.}
    \label{fig:cluster_distribution}
    \phantomsubcaption\label{fig:cluster_distribution_a}
    \phantomsubcaption\label{fig:cluster_distribution_b}
\end{figure*}

\subsubsection{Networking and clustering}
Cesium induces a small degree of 
\ce{BeF4^{2-}}
network disruption, evidenced by a decrease in the simulated network connectivity. \Cref{fig:cluster_distribution} presents the degree of polymerization for each composition and the size distribution of clusters formed by connected \ce{Be-F} polyhedra in our NNMD simulations at \qty{510}{\degree C}. The number of monomeric species increase with the inclusion of Cs, while the polymeric species decrease. The Be cluster size distribution for FLiBe and \ce{FLiBe-CsF} in \cref{fig:cluster_distribution_b} identifies a similar trend with more monomeric species and fewer large clusters with Cs. EPSR models had a similar trend with a decrease in polymerization with the addition of Cs, shown in Fig. S14. 

\begin{figure*}[tb]
    \centering
    \includegraphics[]{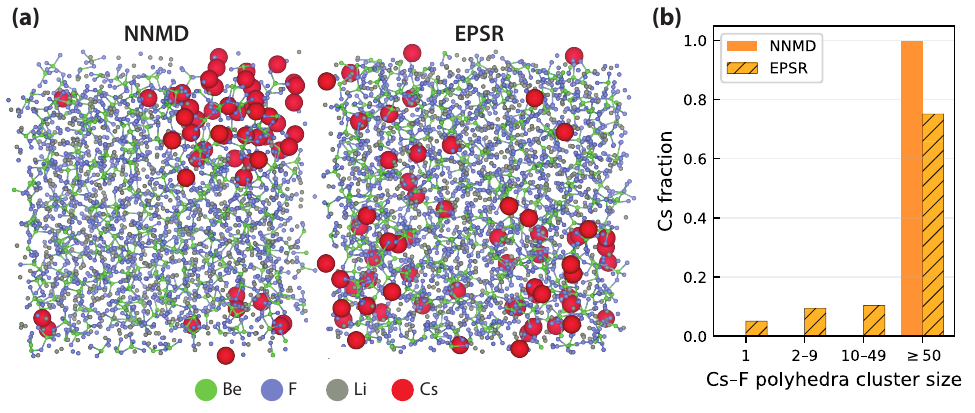}
    \caption{(a) Snapshots of the atomic configuration of \ce{FLiBe-CsF} at \qty{510}{\degree C} from NNMD and EPSR models, and (b) the Cs-F polyhedra cluster size distribution. The snapshot visualization only displays the \ce{Be-F} bonds, highlighting the network formation.}
    \label{fig:CsClustering}
    \phantomsubcaption\label{fig:CsClustering_a}
    \phantomsubcaption\label{fig:CsClustering_b}
\end{figure*}

The cesium fluoride polyhedra mixed with FLiBe clustered together substantially, as seen from both EPSR and NNMD models. Visualization of the atomic configurations are shown by the snapshots in \cref{fig:CsClustering_a}, where Cs atoms are highlighted in red. The cesium clustering was more prominent in the NNMD simulation than the EPSR model. This is evident from the quantitative comparison in \cref{fig:CsClustering_b}, which indicates that most cesium atoms group together into a single cluster in NNMD. The cesium clustering was still large from the EPSR model, but the distribution of cluster sizes is wider than observed by NNMD. The predicted differences in clustering can also be seen in the unweighted \ce{Cs-Cs} partial PDF in Fig. S9. As the observed \ce{Cs-Cs} and \ce{Cs-F} correlations were prominent in the X-ray structure factor, the clustering predicted by EPSR models should be well constrained by the experimental data. This is supported by a better fit in both intensity and position of the EPSR-predicted peak at \qty{1.8}{\per\angstrom}, which corresponds to \ce{Cs-Cs} and \ce{Cs-F} correlations as shown in \cref{fig:csflibe_xray_nnmd_c}. While the \ce{Cs-Cs} clustering is prominent from both simulations, we note that the cesium atoms are not clustered together as metallic particles, as indicated by the CN$_\textrm{RAD}$ in \cref{tab:coords_and_peak} being near zero from both NNMD and ESPR models; instead, the clustering is mediated by \ce{BeF4^{2-}} tetrahedra.

\section{Discussion}\label{sec:discussion}
The \ce{FLiBe-CsF} solid structure was predicted to have \ce{CsBeF3}, \ce{LiF}, and \ce{Li2BeF4} phases near room temperature, according to the thermochemical molten salt database (MSD-TC) \citep{Besmann2021_DevelopingPracticalModels}. The diffraction patterns presented in \cref{fig:xrd_solid_rt_hightemp} did not match the predicted reflections from those phases at room temperature; the \ce{Li2BeF4} phase appeared above \qty{180}{\degree C}. To corroborate these findings, we prepared a powdered \ce{FLiBe-CsF} specimen from the sample batch used for the other measurements in a \qty{1.5}{mm} glass tube for room temperature XRD measurements. Even with the improved powdered specimen combined with translating the X-ray beam across the powdered specimen height, the XRD pattern shown in Fig. S17 matched the result in \cref{fig:xrd_solid_rt_hightemp}. A more comprehensive analysis of the solid structure of \ce{FLiBe-CsF} might reveal evidence for the predicted phases but is beyond the scope of this study. It is important, however, to notice the discrepancy between MSD-TC prediction and experimental results.

In the liquid structure, calculating the CN with the RAD method proved to be superior in separating out the complex coordination environment for each species. With the high concentration opposite-charged pairs \ce{Be-F} and \ce{Li-F}, the GC and RAD methods generally agreed, reinforcing the $r_\textrm{min}$ values chosen for the cutoff distances. More interestingly, the RAD method more accurately separated out the first and second coordination shells for the cesium ions, with the average \ce{Cs-F} CN$_\textrm{RAD}$ being around five. In a \ce{UCl3-NaCl} mixture, cesium was found to have nearly ten-fold coordination with chlorine \citep{Wang2025_CoordinationDrivenMixingBehavior}, matching the value we find with the GC method. However, pure \ce{CsCl} \citep{Wang2025_CoordinationDrivenMixingBehavior} and \ce{CsF} \citep{Matsumiya2002_InvestigationElectricalProperties} were both found to have a five- to six-fold chlorine and fluorine coordination, surprisingly similar to our CN$_\textrm{GC}=5.2$. With the notable amount of cesium clustering and the CsF coordination being similar to pure CsF and CsCl species, the CsF mixed with FLiBe seems to be structured remarkably similarly to pure CsF. Although, the CsF first peak distances are larger in \ce{FLiBe-CsF} because the clustered cesium ions are bridged by \ce{BeF4^{2-}} tetrahedra rather than individual fluorine ions. Additionally, this mixture produces a broader CN distribution and less well-defined polyhedra as shown in \cref{fig:cluster_distribution}.

The EPSR models refined to the experimental structure factor generally agree well with the NNMD simulations with the same trends seen in the CN distributions and bond angles. The \ce{BeF4^{2-}} tetrahedra from EPSR models are less well-defined with a broader angle distribution seen in \cref{fig:CN_dist_bond_angles_b}, causing the EPSR \ce{Be-F} CN$_\textrm{RAD}$ to be higher than NNMD, while the \ce{Be-F} CN$_\textrm{GC}$ values match closely. The largest discrepancy between the two methods can be seen in the \ce{Be-F-Be} ADFs, representing the bridging angle between connected \ce{BeF4^{2-}} polyhedra. Specifically, the EPSR models predict the formation of nearly linear \ce{Be-F-Be} connections, consistent with previously reported EPSR ADFs \citep{Williams_StructureSpeciationMolten}, resulting in a longer \ce{Be-Be} distance than that predicted by NNMD. This is attributed to the neglect of polarization effects, which are particularly important for bivalent cations such as \ce{Be^2+}. Polarization reduces the repulsion between Be ions in connected \ce{BeF4^{2-}} tetrahedra, shortening the \ce{Be-Be} distance and consequently reducing the \ce{Be-F-Be} bond angle. \citep{Rollet2011_StudiesLocalStructures}. NNMD simulations capture polarization and many-body electron effects through training on DFT-predicted forces and energies, leading to a maximum in the \ce{Be-F-Be} ADF at \qty{130}{\degree}, which matches other findings \citep{Winner2021_AbinitioSimulationStudies, Lam2021_ModelingLiFFLiBe, Langford2022_ConstantpotentialMolecularDynamics, Fayfar2024_ComplexStructureMolten, Zakiryanov2026_LiFBeF2MoltenMixtures}. 

The extent of the polymerization between the models differed, with EPSR predicting much longer structures than our NNMD simulations and other studies \citep{Smith2020_NewApproachCoupled}.  The higher fraction of polymeric species from EPSR models, displayed in Fig. S14, can be attributed to the lack of polarization effects in the model. Comparisons between AIMD and rigid ion models (RIM) have found that nonpolarizable RIM simulations predict significantly longer polymers for \ce{ZrF_4} systems \citep{Chahal2022_ShortIntermediateRangeStructure}. Since the only indicators for \ce{BeF4^{2-}} networking in the structure factor are \ce{Be-Be} and \ce{Be-F} correlations, EPSR models do not have enough constraints to accurately predict the networking. As such, the NNMD simulations should result in a more reliable prediction of the network formation due to their ability to account for polarization effects through direct DFT training.

Although the EPSR does not explicitly include polarization effects, the resulting geometric features (e.g., Be-Be distance) could be reproduced through the empirical potential fit to experimental peaks. This, in the case of FLiBe and \ce{FLiBe-CsF}, corresponds to the \ce{Be-Be} partial PDF peak, which makes negligible contributions to the total X-ray and neutron PDFs. As such, the EPSR cannot determine a unique solution for structural features that are effectively unobservable in the measured scattering data; the spatial arrangement and connectivity of the fluoroberyllate network should be interpreted as model-dependent rather than experimentally resolved. 

In contrast, the prominent correlations such as the \ce{Cs-F}, \ce{Cs-Cs}, \ce{Be-F}, \ce{Li-F}, and \ce{F-F} pairs make substantial contribution to the measured structure factor and PDF. Their first peak position and coordination numbers are therefore strongly constrained by the experimental data and can be extracted from the EPSR-derived partial PDFs, which separate the overlapping pair contribution and represent the full peak shapes rather than relying on simple fits to the total-scattering features. Along these lines, the cesium clustering determined by EPSR models was well constrained by the experimental data, resulting in a structure factor more closely aligned with the experiment data in \cref{fig:csflibe_xray_nnmd_c}. The excessive Cs clustering predicted by NNMD simulations could be caused by training on systems with only two Cs atoms. Incorporating many-body \ce{Cs-Cs} interaction and fine-tuning the NNIP on DFT data containing more Cs atoms can provide a more realistic model to examine cluster. This investigation will be left for future work. 

\section{Conclusions}\label{sec:conclusion}
The structure of FLiBe with 5 mol\% CsF was investigated using experimental X-ray and neutron diffraction along with NNMD simulations and EPSR models fit to the experimental data. These models enable structural features to be quantified such as the first peak distances, coordination numbers, networking, and clustering. Cesium is an important fission product in salt-fueled reactors due to its high yield and long-lived half lives. Characterizing the structural variation due to fission products will be essential for predicting the evolution of thermophysical properties during reactor operation. In the solid structure, the \ce{FLiBe-CsF} XRD pattern did not match predicted reflections from \ce{Li2BeF4}, \ce{CsBeF3}, or \ce{LiF} phases. Upon heating, reflections from the FLiBe phase appeared near \qty{180}{\degree C}, and melting completed around \qty{450}{\degree C}. In the molten structure, the new Cs-bearing correlations accounted for the major changes in the total structure factor and PDF, with minimal changes occurring to the FLiBe correlations. 
Cesium perturbed the local FLiBe structure to a small degree, with only minor changes in the first peak distances, CNs, and bond angles. Cesium slightly disrupted the fluoroberyllate network, increasing the number of monomers and decreasing the number of polymers. The extent of the fluoroberyllate polymerization was not directly resolvable from experimental scattering measurements, leading to model-dependent findings. NNMD simulations more accurately predict the amount of polymerization due to the polarization effects necessary to accurately model multivalent cation species. When mixed with FLiBe, the cesium atoms clustered together considerably, with intermediary \ce{BeF4^{2-}} tetrahedra bridging the cesium atoms. The EPSR models were well-constrained by the experimental cesium correlations resulting in EPSR more accurately modeling cesium clustering than NNMD. Together, this study provided an experimental benchmark for atomistic simulations, distinguish experimentally constrained structure features from model-dependent predictions, and improved confidence in models used to predict structural evolution under reactor-relevant conditions. 

\section{Acknowledgments}
This material is based upon work supported by the U.S. Department of Energy,  under Award Number DE-NE0009486 (measurements and data analysis) and DE-NE0008751 (specimens and data collection). A portion of this research used resources at Spallation Neutron Source, a DOE Office of Science User Facility operated by the Oak Ridge National Laboratory. The beam time was allocated to NOMAD on proposal number IPTS-29309.1 and 28304.1.
This research used beamline 28-ID-1 (PDF) of the National Synchrotron Light Source II, a U.S. Department of Energy (DOE) Office of Science User Facility operated for the DOE Office of Science by Brookhaven National Laboratory under Contract No. DE-SC0012704. This research used resources of the National Energy Research Scientific Computing Center, a DOE Office of Science User Facility supported by the Office of Science of the U.S. Department of Energy under Contract No. DE-AC02-05CH11231 using NERSC award DDR-ERCAP0038149.

We acknowledge discussions with Prof. Juliano Schorne Pinto (University of South Carolina) on the phase diagram of \ce{FLiBe-CsF}.

\bibliography{References}

@article{Alderman2015_LiquidB2O31700,
  title = {Liquid {{B2O3}} up to 1700 {{K}}: X-Ray Diffraction and Boroxol Ring Dissolution},
  shorttitle = {Liquid {{B2O3}} up to 1700 {{K}}},
  author = {Alderman, O. L. G. and Ferlat, G. and Baroni, A. and Salanne, M. and Micoulaut, M. and Benmore, C. J. and Lin, A. and Tamalonis, A. and Weber, J. K. R.},
  year = 2015,
  month = oct,
  journal = {Journal of Physics: Condensed Matter},
  volume = {27},
  number = {45},
  pages = {455104},
  publisher = {IOP Publishing},
  issn = {0953-8984},
  doi = {10.1088/0953-8984/27/45/455104},
  urldate = {2023-01-05},
  langid = {english}
}

@article{Andreades2016_DesignSummaryMarkI,
  title = {Design {{Summary}} of the {{Mark-I Pebble-Bed}}, {{Fluoride Salt}}--{{Cooled}}, {{High-Temperature Reactor Commercial Power Plant}}},
  author = {Andreades, Charalampos and Cisneros, Anselmo T. and Choi, Jae Keun and Chong, Alexandre Y. K. and Fratoni, Massimiliano and Hong, Sea and Huddar, Lakshana R. and Huff, Kathryn D. and Kendrick, James and Krumwiede, David L. and Laufer, Michael R. and Munk, Madicken and Scarlat, Raluca O. and Zweibau, Nicolas},
  year = 2016,
  month = sep,
  journal = {Nuclear Technology},
  volume = {195},
  number = {3},
  pages = {223--238},
  publisher = {Taylor \& Francis},
  issn = {0029-5450},
  doi = {10.13182/NT16-2},
  urldate = {2023-03-27}
}

@article{Attarian2024_StudiesNiCrComplexation,
  title = {Studies of {{Ni-Cr}} Complexation in {{FLiBe}} Molten Salt Using Machine Learning Interatomic Potentials},
  author = {Attarian, Siamak and Morgan, Dane and Szlufarska, Izabela},
  year = 2024,
  month = apr,
  journal = {Journal of Molecular Liquids},
  volume = {400},
  pages = {124521},
  issn = {0167-7322},
  doi = {10.1016/j.molliq.2024.124521},
  urldate = {2026-06-04}
}

@article{Baes1970_PolymerModelBeF2,
  title = {A Polymer Model for {{BeF2}} and {{SiO2}} Melts},
  author = {Baes, C. F.},
  year = 1970,
  month = jan,
  journal = {Journal of Solid State Chemistry},
  volume = {1},
  number = {2},
  pages = {159--169},
  issn = {0022-4596},
  doi = {10.1016/0022-4596(70)90008-3},
  urldate = {2023-06-20},
  langid = {english}
}

@article{Baral2021_TemperatureDependentPropertiesMolten,
  title = {Temperature-{{Dependent Properties}} of {{Molten Li2BeF4 Salt Using Ab Initio Molecular Dynamics}}},
  author = {Baral, Khagendra and San, Saro and Sakidja, Ridwan and Couet, Adrien and Sridharan, Kumar and Ching, Wai-Yim},
  year = 2021,
  month = aug,
  journal = {ACS Omega},
  volume = {6},
  number = {30},
  pages = {19822--19835},
  publisher = {American Chemical Society},
  doi = {10.1021/acsomega.1c02528},
  urldate = {2023-03-16}
}

@article{Benes2021_CesiumIodineRelease,
  title = {Cesium and Iodine Release from Fluoride-Based Molten Salt Reactor Fuel},
  author = {Bene{\v s}, O. and Capelli, E. and Morelov{\'a}, N. and Colle, J.-Y. and Tosolin, A. and Wiss, T. and Cremer, B. and Konings, R. J. M.},
  year = 2021,
  month = apr,
  journal = {Physical Chemistry Chemical Physics},
  volume = {23},
  number = {15},
  pages = {9512--9523},
  publisher = {The Royal Society of Chemistry},
  issn = {1463-9084},
  doi = {10.1039/D0CP05794K},
  urldate = {2022-06-27},
  langid = {english}
}

@article{Besmann2021_DevelopingPracticalModels,
  title = {Developing {{Practical Models}} of {{Complex Salts}} for {{Molten Salt Reactors}}},
  author = {Besmann, Theodore M. and {Schorne-Pinto}, Juliano},
  year = 2021,
  month = sep,
  journal = {Thermo},
  volume = {1},
  number = {2},
  pages = {168--178},
  publisher = {Multidisciplinary Digital Publishing Institute},
  issn = {2673-7264},
  doi = {10.3390/thermo1020012},
  urldate = {2024-03-12},
  copyright = {http://creativecommons.org/licenses/by/3.0/},
  langid = {english}
}

@article{Bettis1957_AircraftReactorExperiment,
  title = {The {{Aircraft Reactor Experiment}}---{{Operation}}},
  author = {Bettis, E. S. and Cottrell, W. B. and Mann, E. R. and Meem, J. L. and Whitman, G. D.},
  year = 1957,
  month = nov,
  journal = {Nuclear Science and Engineering},
  volume = {2},
  number = {6},
  pages = {841--853},
  issn = {0029-5639},
  doi = {10.13182/NSE57-A35497}
}

@article{Cantor1969_ViscosityDensityMolten,
  title = {Viscosity and {{Density}} in {{Molten BeF2}}--{{LiF Solutions}}},
  author = {Cantor, S. and Ward, W. T. and Moynihan, C. T.},
  year = 1969,
  month = apr,
  journal = {The Journal of Chemical Physics},
  volume = {50},
  number = {7},
  pages = {2874--2879},
  publisher = {American Institute of Physics},
  issn = {0021-9606},
  doi = {10.1063/1.1671478},
  urldate = {2022-12-12}
}

@article{Carlson2021_MoltenSaltReactors,
  title = {Molten Salt Reactors and Electrochemical Reprocessing: Synthesis and Chemical Durability of Potential Waste Forms for Metal and Salt Waste Streams},
  shorttitle = {Molten Salt Reactors and Electrochemical Reprocessing},
  author = {Carlson, Krista and Gardner, Levi and Moon, Jeremy and Riley, Brian and Amoroso, Jake and Chidambaram, Dev},
  year = 2021,
  month = jul,
  journal = {International Materials Reviews},
  volume = {66},
  number = {5},
  pages = {339--363},
  publisher = {SAGE Publications},
  issn = {0950-6608},
  doi = {10.1080/09506608.2020.1801229},
  urldate = {2026-03-04},
  langid = {english}
}

@article{Chahal2022_ShortIntermediateRangeStructure,
  title = {Short- to {{Intermediate-Range Structure}}, {{Transport}}, and {{Thermophysical Properties}} of {{LiF}}--{{NaF}}--{{ZrF4 Molten Salts}}},
  author = {Chahal, Rajni and Banerjee, Shubhojit and Lam, Stephen T.},
  year = 2022,
  month = mar,
  journal = {Frontiers in Physics},
  volume = {10},
  pages = {830468},
  issn = {2296-424X},
  doi = {10.3389/fphy.2022.830468},
  urldate = {2023-12-12}
}

@article{Chahal2022_TransferableDeepLearning,
  title = {Transferable {{Deep Learning Potential Reveals Intermediate-Range Ordering Effects}} in {{LiF}}--{{NaF}}--{{ZrF4 Molten Salt}}},
  author = {Chahal, Rajni and Roy, Santanu and Brehm, Martin and Banerjee, Shubhojit and Bryantsev, Vyacheslav and Lam, Stephen T.},
  year = 2022,
  month = dec,
  journal = {JACS Au},
  volume = {2},
  number = {12},
  pages = {2693--2702},
  publisher = {American Chemical Society},
  doi = {10.1021/jacsau.2c00526},
  urldate = {2023-01-17}
}

@article{Clark2021_ComplexationMoFLiNaK,
  title = {Complexation of {{Mo}} in {{FLiNaK Molten Salt}}: {{Insight}} from {{Ab Initio Molecular Dynamics}}},
  shorttitle = {Complexation of {{Mo}} in {{FLiNaK Molten Salt}}},
  author = {Clark, Austin David and Lee, Wan Luan and Solano, Andrew Russell and Williams, Tyler Bruce and Meyer, Gabriel Scott and Tait, Granite J. and Battraw, Ben C. and Nickerson, Stella D.},
  year = 2021,
  month = jan,
  journal = {The Journal of Physical Chemistry B},
  volume = {125},
  number = {1},
  pages = {211--218},
  publisher = {American Chemical Society},
  issn = {1520-6106},
  doi = {10.1021/acs.jpcb.0c07354},
  urldate = {2026-06-04}
}

@techreport{Compere1975_FissionProductBehavior,
  title = {Fission Product Behavior in the {{Molten Salt Reactor Experiment}}},
  author = {Compere, E and Kirslis, S and Bohlmann, E and Blankenship, F and Grimes, W},
  year = 1975,
  month = oct,
  number = {ORNL--4865, 4077644},
  pages = {ORNL--4865, 4077644},
  doi = {10.2172/4077644},
  urldate = {2023-03-15},
  langid = {english}
}

@article{Dai2018_MolecularDynamicsInvestigation,
  title = {Molecular Dynamics Investigation on the Local Structures and Transport Properties of Uranium Ion in {{LiCl-KCl}} Molten Salt},
  author = {Dai, Jian-Xing and Zhang, Wei and Ren, Cui-Lan and Han, Han and Guo, Xiao-Jing and Li, Qing-Nuan},
  year = 2018,
  month = dec,
  journal = {Journal of Nuclear Materials},
  series = {Special {{Section}} on "18th {{International Conference}} on {{Fusion Reactor Materials}}"},
  volume = {511},
  pages = {75--82},
  issn = {0022-3115},
  doi = {10.1016/j.jnucmat.2018.08.052},
  urldate = {2026-06-04}
}

@techreport{DeVan1962_CorrosionBehaviorReactor,
  title = {Corrosion {{Behavior}} of {{Reactor Materials}} in {{Fluoride Salt Mixtures}}},
  author = {DeVan, J. H. and Evans, I. I. I.},
  year = 1962,
  month = sep,
  number = {ORNL-TM-328},
  institution = {Oak Ridge National Lab. (ORNL), Oak Ridge, TN (United States)},
  doi = {10.2172/4774669},
  urldate = {2023-03-15},
  langid = {english}
}

@article{Du2009_MolecularDynamicsSimulation,
  title = {A Molecular Dynamics Simulation Interpretation of Neutron and X-Ray Diffraction Measurements on Single Phase {{Y2O3}}--{{Al2O3}} Glasses},
  author = {Du, Jincheng and Benmore, Chris J. and Corrales, Rene and Hart, Robert T. and Weber, J. K. Richard},
  year = 2009,
  month = apr,
  journal = {Journal of Physics: Condensed Matter},
  volume = {21},
  number = {20},
  pages = {205102},
  issn = {0953-8984},
  doi = {10.1088/0953-8984/21/20/205102},
  urldate = {2022-11-16},
  langid = {english}
}

@article{Fayfar2023_InSituAnalysisCorrosion,
  title = {In-{{Situ Analysis}} of {{Corrosion Products}} in {{Molten Salt}}: {{X-ray Absorption Reveals Both Ionic}} and {{Metallic Species}}},
  shorttitle = {In-{{Situ Analysis}} of {{Corrosion Products}} in {{Molten Salt}}},
  author = {Fayfar, Sean and Zheng, Guiqiu and Sprouster, David and Marshall, Matthew S. J. and Stavitski, Eli and Leshchev, Denis and Khaykovich, Boris},
  year = 2023,
  month = jul,
  journal = {ACS Omega},
  volume = {8},
  number = {27},
  pages = {24673--24679},
  publisher = {American Chemical Society},
  doi = {10.1021/acsomega.3c03448},
  urldate = {2023-07-11},
  copyright = {All rights reserved}
}

@article{Fayfar2024_ComplexStructureMolten,
  title = {Complex {{Structure}} of {{Molten FLiBe}} (2\$\textbackslash mathrm\textbraceleft{{Li}}\textbraceright\textbackslash mathrm\textbraceleft{{F}}\textbraceright\$--\$\textbackslash mathrm\textbraceleft{{Be}}\textbraceright\textbackslash mathrm\textbraceleft{{F}}\textbraceright\$\$\textbraceleft\textbraceright\_\textbraceleft 2\textbraceright\$) {{Examined}} by {{Experimental Neutron Scattering}}, {{X-Ray Scattering}}, and {{Deep-Neural-Network Based Molecular Dynamics}}},
  author = {Fayfar, Sean and Chahal, Rajni and Williams, Haley and Gardner, D. Nathanael and Zheng, Guiqiu and Sprouster, David and Neuefeind, J{\"o}rg C. and Olds, Dan and Hwang, Andrea and Mcfarlane, Joanna and Gallagher, Ryan C. and Asta, Mark and Lam, Stephen and Scarlat, Raluca O. and Khaykovich, Boris},
  year = 2024,
  month = jan,
  journal = {PRX Energy},
  volume = {3},
  number = {1},
  pages = {013001},
  publisher = {American Physical Society},
  doi = {10.1103/PRXEnergy.3.013001},
  urldate = {2024-01-05},
  copyright = {All rights reserved}
}

@article{Fischer2006_NeutronXrayDiffraction,
  title = {Neutron and X-Ray Diffraction Studies of Liquids and Glasses},
  author = {Fischer, Henry E. and Barnes, Adrian C. and Salmon, Philip S.},
  year = 2006,
  journal = {Reports on Progress in Physics},
  volume = {69},
  number = {1},
  pages = {233--299},
  issn = {00344885},
  doi = {10.1088/0034-4885/69/1/R05}
}

@article{Forsberg2020_FusionBlanketsFluoridesaltcooled,
  title = {Fusion Blankets and Fluoride-Salt-Cooled High-Temperature Reactors with Flibe Salt Coolant: Common Challenges, Tritium Control, and Opportunities for Synergistic Development Strategies between Fission, Fusion, and Solar Salt Technologies},
  shorttitle = {Fusion Blankets and Fluoride-Salt-Cooled High-Temperature Reactors with Flibe Salt Coolant},
  author = {Forsberg, Charles and Zheng, Guiqiu and Ballinger, Ronald G. and Lam, Stephen T.},
  year = 2020,
  journal = {Nuclear Technology},
  volume = {206},
  number = {11},
  pages = {1778--1801},
  publisher = {Taylor \& Francis}
}

@article{Frandsen2020_StructureMoltenFLiNaK,
  title = {The Structure of Molten {{FLiNaK}}},
  author = {Frandsen, Benjamin A. and Nickerson, Stella D. and Clark, Austin D. and Solano, Andrew and Baral, Raju and Williams, Johnny and Neuefeind, J{\"o}rg and Memmott, Matthew},
  year = 2020,
  month = aug,
  journal = {Journal of Nuclear Materials},
  volume = {537},
  pages = {152219},
  issn = {0022-3115},
  doi = {10.1016/j.jnucmat.2020.152219},
  urldate = {2022-06-27},
  langid = {english}
}

@article{Gallington2023_ReviewCurrentSoftware,
  title = {Review of {{Current Software}} for {{Analyzing Total X-ray Scattering Data}} from {{Liquids}}},
  author = {Gallington, Leighanne C. and Wilke, Stephen K. and Kohara, Shinji and Benmore, Chris J.},
  year = 2023,
  month = jun,
  journal = {Quantum Beam Science},
  volume = {7},
  number = {2},
  pages = {20},
  publisher = {Multidisciplinary Digital Publishing Institute},
  issn = {2412-382X},
  doi = {10.3390/qubs7020020},
  urldate = {2023-06-29},
  copyright = {http://creativecommons.org/licenses/by/3.0/},
  langid = {english}
}

@article{Grimme2010_ConsistentAccurateInitio,
  title = {A Consistent and Accurate Ab Initio Parametrization of Density Functional Dispersion Correction ({{DFT-D}}) for the 94 Elements {{H-Pu}}},
  author = {Grimme, Stefan and Antony, Jens and Ehrlich, Stephan and Krieg, Helge},
  year = 2010,
  month = apr,
  journal = {The Journal of Chemical Physics},
  volume = {132},
  number = {15},
  pages = {154104},
  publisher = {American Institute of Physics},
  issn = {0021-9606},
  doi = {10.1063/1.3382344},
  urldate = {2022-12-02}
}

@article{Guo2023_XrayMolecularDynamics,
  title = {X-Ray and Molecular Dynamics Study of the Temperature-Dependent Structure of {{FLiNaK}}},
  author = {Guo, Jicheng and Zhang, Yifan and Ludwig, Karl and Yan, Haoxuan and Levy, Alexander and Wadehra, Anubhav and Gao, Michael C. and Benmore, Chris and Rose, Melissa and Condon, Nicholas and Powell, Adam and Zhong, Yu and Pal, Uday},
  year = 2023,
  month = dec,
  journal = {Nuclear Materials and Energy},
  volume = {37},
  pages = {101530},
  issn = {2352-1791},
  doi = {10.1016/j.nme.2023.101530},
  urldate = {2024-01-12}
}

@article{Heaton2006_FirstPrinciplesDescriptionLiquid,
  title = {A {{First-Principles Description}} of {{Liquid BeF2}} and {{Its Mixtures}} with {{LiF}}:\, 1. {{Potential Development}} and {{Pure BeF2}}},
  shorttitle = {A {{First-Principles Description}} of {{Liquid BeF2}} and {{Its Mixtures}} with {{LiF}}},
  author = {Heaton, Robert J. and Brookes, Richard and Madden, Paul A. and Salanne, Mathieu and Simon, Christian and Turq, Pierre},
  year = 2006,
  month = jun,
  journal = {The Journal of Physical Chemistry B},
  volume = {110},
  number = {23},
  pages = {11454--11460},
  publisher = {American Chemical Society},
  issn = {1520-6106},
  doi = {10.1021/jp061000+},
  urldate = {2023-09-11}
}

@article{Heinen2022_LiquidDiffractSoftwareLiquid,
  title = {{{LiquidDiffract}}: Software for Liquid Total Scattering Analysis},
  shorttitle = {{{LiquidDiffract}}},
  author = {Heinen, Benedict J. and Drewitt, James W. E.},
  year = 2022,
  month = may,
  journal = {Physics and Chemistry of Minerals},
  volume = {49},
  number = {5},
  pages = {9},
  issn = {1432-2021},
  doi = {10.1007/s00269-022-01186-6},
  urldate = {2023-05-09},
  langid = {english}
}

@article{Higham2016_LocallyAdaptiveMethod,
  title = {Locally Adaptive Method to Define Coordination Shell},
  author = {Higham, Jonathan and Henchman, Richard H.},
  year = 2016,
  month = aug,
  journal = {The Journal of Chemical Physics},
  volume = {145},
  number = {8},
  pages = {084108},
  issn = {0021-9606},
  doi = {10.1063/1.4961439},
  urldate = {2026-06-10}
}

@article{Higham2018_OvercomingLimitationsCutoffs,
  title = {Overcoming the Limitations of Cutoffs for Defining Atomic Coordination in Multicomponent Systems},
  author = {Higham, Jonathan and Henchman, Richard H.},
  year = 2018,
  journal = {Journal of Computational Chemistry},
  volume = {39},
  number = {12},
  pages = {705--710},
  issn = {1096-987X},
  doi = {10.1002/jcc.25137},
  urldate = {2026-05-04},
  copyright = {\copyright{} 2017 Wiley Periodicals, Inc.},
  langid = {english}
}

@article{Hoover1985_CanonicalDynamicsEquilibrium,
  title = {Canonical Dynamics: {{Equilibrium}} Phase-Space Distributions},
  shorttitle = {Canonical Dynamics},
  author = {Hoover, William G.},
  year = 1985,
  month = mar,
  journal = {Physical Review A},
  volume = {31},
  number = {3},
  pages = {1695--1697},
  publisher = {American Physical Society},
  doi = {10.1103/PhysRevA.31.1695},
  urldate = {2022-12-02}
}

@article{Hunsberger2026_AcidbaseEquilibriaMolten,
  title = {Acid-Base Equilibria in Molten Fluoride Salts and Their Impact on Molten Salt Reactors},
  author = {Hunsberger, O. and Rakib, G. S. and Vergari, L.},
  year = 2026,
  month = apr,
  journal = {Journal of Fluorine Chemistry},
  pages = {110569},
  issn = {0022-1139},
  doi = {10.1016/j.jfluchem.2026.110569},
  urldate = {2026-04-16}
}

@article{Ito2005_AdvancesMoltenSalt,
  title = {Advances in {{Molten Salt Electrochemistry}} towards {{Future Energy Systems}}},
  author = {Ito, Yasuhiko},
  year = 2005,
  journal = {Electrochemistry},
  volume = {73},
  number = {8},
  pages = {545--551},
  doi = {10.5796/electrochemistry.73.545}
}

@inproceedings{Ivashkevych2020_HardXRayPair,
  title = {Hard {{X-Ray Pair Distribution Function}} ({{PDF}}) {{Beamline}} and {{End-Station Control System}}},
  booktitle = {17th {{International Conference}} on {{Accelerator}} and {{Large Experimental Physics Control Systems}} ({{ICALEPCS}}'19), {{New York}}, {{NY}}, {{USA}}, 05-11 {{October}} 2019},
  author = {Ivashkevych, Oksana and Abeykoon, Milinda and Adams, Julian and Bischof, Garrett and Dooryhee, Eric and Li, Ji and Petkus, Robert and Trunk, John and Yin, Zhijian},
  year = 2020,
  month = aug,
  pages = {1584--1589},
  publisher = {JACOW Publishing, Geneva, Switzerland},
  issn = {2226-0358},
  doi = {10.18429/JACoW-ICALEPCS2019-THBPP04},
  urldate = {2023-03-31},
  isbn = {978-3-95450-209-7},
  langid = {english}
}

@book{Janz_MoltenSaltsHandbook,
  title = {Molten {{Salts Handbook}}},
  author = {Janz, George J.},
  year = 1967,
  publisher = {Academic Press},
  doi = {10.1016/B978-0-123-95642-2.X5001-1},
  isbn = {978-0-12-395642-2},
  langid = {english},
  lccn = {546.34}
}

@article{Jiang2026_ConcentrationTransferableDeep,
  title = {Concentration--Transferable Deep Potential Molecular Dynamics: Unraveling Component--Structure--Transport in Molten {{LiF}}--{{BeF2}}--{{EuF2}}},
  shorttitle = {Concentration--Transferable Deep Potential Molecular Dynamics},
  author = {Jiang, Yuanyuan and Li, Xuejiao and Gong, Yu},
  year = 2026,
  month = jun,
  journal = {Physical Chemistry Chemical Physics},
  publisher = {The Royal Society of Chemistry},
  issn = {1463-9084},
  doi = {10.1039/D6CP01507G},
  urldate = {2026-06-25},
  langid = {english}
}

@article{Kieffer2013_PyFAIVersatileLibrary,
  title = {{{PyFAI}}, a Versatile Library for Azimuthal Regrouping},
  author = {Kieffer, J{\'e}r{\^o}me and Karkoulis, Dimitrios},
  year = 2013,
  month = mar,
  journal = {Journal of Physics: Conference Series},
  volume = {425},
  number = {20},
  pages = {202012},
  issn = {1742-6596},
  doi = {10.1088/1742-6596/425/20/202012},
  urldate = {2022-10-28},
  langid = {english}
}

@techreport{Koger1972_EvaluationHastelloyAlloys,
  title = {Evaluation of {{Hastelloy N}} Alloys after Nine Years Exposure to Both a Molten Fluoride Salt and Air at Temperatures from 700 to 560 {$^\circ$}{{C}}},
  author = {Koger, J. W.},
  year = 1972,
  month = dec,
  number = {ORNL-TM-4189},
  address = {United States},
  doi = {10.2172/4468052},
  langid = {english}
}

@techreport{Koger1973_CorrosionProductDeposition,
  title = {Corrosion Product Deposition in Molten Fluoride Salt Systems.},
  author = {Koger, J. W.},
  year = 1973,
  month = apr,
  number = {CONF-730308-1},
  institution = {Oak Ridge National Lab. (ORNL), Oak Ridge, TN (United States)},
  doi = {10.2172/4567834},
  urldate = {2023-03-15},
  langid = {english}
}

@article{Kresse1996_EfficientIterativeSchemes,
  title = {Efficient Iterative Schemes for Ab Initio Total-Energy Calculations Using a Plane-Wave Basis Set},
  author = {Kresse, G. and Furthm{\"u}ller, J.},
  year = 1996,
  month = oct,
  journal = {Physical Review B},
  volume = {54},
  number = {16},
  pages = {11169--11186},
  publisher = {American Physical Society},
  doi = {10.1103/PhysRevB.54.11169},
  urldate = {2022-12-02}
}

@article{Lam2021_ModelingLiFFLiBe,
  title = {Modeling {{LiF}} and {{FLiBe Molten Salts}} with {{Robust Neural Network Interatomic Potential}}},
  author = {Lam, Stephen T. and Li, Qing-Jie and Ballinger, Ronald and Forsberg, Charles and Li, Ju},
  year = 2021,
  month = jun,
  journal = {ACS Applied Materials \& Interfaces},
  volume = {13},
  number = {21},
  pages = {24582--24592},
  publisher = {American Chemical Society},
  issn = {1944-8244},
  doi = {10.1021/acsami.1c00604},
  urldate = {2023-04-05}
}

@article{Langford2022_ConstantpotentialMolecularDynamics,
  title = {Constant-Potential Molecular Dynamics Simulations of Molten Salt Double Layers for {{FLiBe}} and {{FLiNaK}}},
  author = {Langford, Luke and Winner, Nicholas and Hwang, Andrea and Williams, Haley and Vergari, Lorenzo and Scarlat, Raluca O. and Asta, Mark},
  year = 2022,
  month = sep,
  journal = {The Journal of Chemical Physics},
  volume = {157},
  number = {9},
  pages = {094705},
  publisher = {American Institute of Physics},
  issn = {0021-9606},
  doi = {10.1063/5.0097697},
  urldate = {2023-01-23}
}

@article{Li2021_ComplexStructureMoltena,
  title = {Complex {{Structure}} of {{Molten NaCl}}--{{CrCl}} 3 {{Salt}}: {{Cr}}--{{Cl Octahedral Network}} and {{Intermediate-Range Order}}},
  author = {Li, Qing-Jie and Sprouster, David and Zheng, Guiqiu and Neuefeind, J{\"o}rg C and Braatz, Alexander D and Mcfarlane, Joanna and Olds, Daniel and Lam, Stephen and Li, Ju and Khaykovich, Boris},
  year = 2021,
  month = apr,
  journal = {ACS Applied Energy Materials},
  volume = {4},
  number = {4},
  pages = {3044--3056},
  publisher = {American Chemical Society},
  issn = {2574-0962},
  doi = {10.1021/acsaem.0c02678}
}

@article{Li2024_CompositionalTransferabilityDeep,
  title = {Compositional Transferability of Deep Potential in Molten {{LiF}}--{{BeF2}} and {{LaF3}} Mixtures: Prediction of Density, Viscosity, and Local Structure},
  shorttitle = {Compositional Transferability of Deep Potential in Molten {{LiF}}--{{BeF2}} and {{LaF3}} Mixtures},
  author = {Li, Xuejiao and Xu, Tingrui and Gong, Yu},
  year = 2024,
  month = apr,
  journal = {Physical Chemistry Chemical Physics},
  volume = {26},
  number = {15},
  pages = {12044--12052},
  publisher = {The Royal Society of Chemistry},
  issn = {1463-9084},
  doi = {10.1039/D4CP00079J},
  urldate = {2026-06-02},
  langid = {english}
}

@article{Li2025_ChemicalFootprintsThorium,
  title = {Chemical {{Footprints}} of {{Thorium}} and {{Uranium}} in {{Molten LiF}}--{{BeF2 Explored}} by {{First-Principles Molecular Dynamics Simulations}}},
  author = {Li, Xuejiao and Jiang, Yuanyuan and Wang, Yuanyuan and Cui, Shiqiang and Gong, Yu},
  year = 2025,
  month = feb,
  journal = {The Journal of Physical Chemistry A},
  volume = {129},
  number = {6},
  pages = {1583--1590},
  issn = {1089-5639},
  doi = {10.1021/acs.jpca.4c07084},
  urldate = {2026-08-21}
}

@article{Li2026_AtomicScaleEuropiumSolute,
  title = {Atomic-{{Scale Europium Solute Landscape Governs}} the {{Local Structure}} and {{Transport Properties}} of the {{Molten FLiBe Salt}}},
  author = {Li, Xuejiao and Jiang, Yuanyuan and Gong, Yu},
  year = 2026,
  month = feb,
  journal = {The Journal of Physical Chemistry B},
  volume = {130},
  number = {9},
  pages = {2597--2604},
  issn = {1520-6106},
  doi = {10.1021/acs.jpcb.5c07601},
  urldate = {2026-07-29}
}

@article{MacPherson1985_MoltenSaltReactor,
  title = {The {{Molten Salt Reactor Adventure}}},
  author = {MacPherson, H. G.},
  year = 1985,
  month = aug,
  journal = {Nuclear Science and Engineering},
  volume = {90},
  number = {4},
  pages = {374--380},
  publisher = {Taylor \& Francis},
  issn = {0029-5639},
  doi = {10.13182/NSE90-374},
  urldate = {2022-07-14}
}

@article{Martinez2009_PACKMOLPackageBuilding,
  title = {{PACKMOL: A package for building initial configurations for molecular dynamics simulations}},
  shorttitle = {{PACKMOL}},
  author = {Mart{\'i}nez, L. and Andrade, R. and Birgin, E. G. and Mart{\'i}nez, J. M.},
  year = 2009,
  journal = {Journal of Computational Chemistry},
  volume = {30},
  number = {13},
  pages = {2157--2164},
  issn = {1096-987X},
  doi = {10.1002/jcc.21224},
  urldate = {2022-12-02},
  langid = {ngerman}
}

@article{Matsumiya2002_InvestigationElectricalProperties,
  title = {Investigation on the Electrical Properties of Molten Quaternary Systems ({{Li}}, {{Na}}, {{K}}, {{Cs}}){{Cl}} and ({{Li}}, {{Na}}, {{K}}, {{Cs}}){{F}} by {{MD}} Simulation},
  author = {Matsumiya, Masahiko and Shin, Woosuck and Izu, Noriya and Murayama, Norimitsu},
  year = 2002,
  month = jun,
  journal = {Journal of Electroanalytical Chemistry},
  volume = {528},
  number = {1-2},
  pages = {103--113},
  issn = {15726657},
  doi = {10.1016/S0022-0728(02)00895-1},
  urldate = {2026-08-27},
  copyright = {https://www.elsevier.com/tdm/userlicense/1.0/},
  langid = {english}
}

@article{McDonnell2017_ADDIEADvancedDIffraction,
  title = {{{ADDIE}}: {{ADvanced DIffraction Environment}} -- a Software Environment for Analyzing Neutron Diffraction Data},
  shorttitle = {{{ADDIE}}},
  author = {McDonnell, M. T. and Olds, D. P. and Page, K. L. and Neufeind, J. C. and Tucker, M. G. and Bilheux, J. C. and Zhou, W. and Peterson, P. F.},
  year = 2017,
  month = may,
  journal = {Acta Crystallographica Section A Foundations and Advances},
  volume = {73},
  number = {a1},
  pages = {a377-a377},
  issn = {2053-2733},
  doi = {10.1107/S0108767317096325},
  urldate = {2022-12-16},
  langid = {english}
}

@article{Merk2018_DemandDrivenSalt,
  title = {Demand Driven Salt Clean-up in a Molten Salt Fast Reactor -- {{Defining}} a Priority List},
  author = {Merk, B. and Litskevich, D. and Gregg, R. and Mount, A. R.},
  year = 2018,
  month = mar,
  journal = {PLOS ONE},
  volume = {13},
  number = {3},
  pages = {e0192020},
  publisher = {Public Library of Science},
  issn = {1932-6203},
  doi = {10.1371/journal.pone.0192020},
  urldate = {2026-04-28},
  langid = {english}
}

@article{Metropolis1953_EquationStateCalculations,
  title = {Equation of {{State Calculations}} by {{Fast Computing Machines}}},
  author = {Metropolis, Nicholas and Rosenbluth, Arianna W. and Rosenbluth, Marshall N. and Teller, Augusta H. and Teller, Edward},
  year = 1953,
  month = jun,
  journal = {The Journal of Chemical Physics},
  volume = {21},
  number = {6},
  pages = {1087--1092},
  issn = {0021-9606},
  doi = {10.1063/1.1699114},
  urldate = {2026-06-04}
}

@article{Moon2024_DensityMeasurementsMolten,
  title = {Density {{Measurements}} of {{Molten LiF}}--{{BeF2}} and {{LiF}}--{{BeF2}}--{{LaF3 Salt Mixtures}} by {{Neutron Radiography}}},
  author = {Moon, Jisue and McFarlane, Joanna and Andrews, Hunter B. and Robb, Kevin R. and Ross, Molly and Sulejmanovic, Dino and Zhang, Yuxuan and Stringfellow, Erik and Agca, Can and {Schorne-Pinto}, Juliano and Besmann, Theodore M.},
  year = 2024,
  month = jun,
  journal = {ACS Omega},
  volume = {9},
  number = {25},
  pages = {27204--27213},
  publisher = {American Chemical Society},
  doi = {10.1021/acsomega.4c01446},
  urldate = {2026-02-20}
}

@article{Nam2014_FirstprinciplesMolecularDynamics,
  title = {First-Principles Molecular Dynamics Modeling of the Molten Fluoride Salt with {{Cr}} Solute},
  author = {Nam, H. O. and Bengtson, A. and V{\"o}rtler, K. and Saha, S. and Sakidja, R. and Morgan, D.},
  year = 2014,
  journal = {Journal of Nuclear Materials},
  volume = {449},
  number = {1-3},
  pages = {148--157},
  publisher = {Elsevier B.V.},
  issn = {00223115},
  doi = {10.1016/j.jnucmat.2014.03.014}
}

@article{Neuefeind2002_HighEnergyXRD,
  title = {High Energy {{XRD}} Investigations of Liquids},
  author = {Neuefeind, J.},
  year = 2002,
  month = may,
  journal = {Journal of Molecular Liquids},
  series = {Molecular {{Structure}} and {{Dynamics}} in {{Liquids}}},
  volume = {98--99},
  pages = {87--95},
  issn = {0167-7322},
  doi = {10.1016/S0167-7322(01)00312-9},
  urldate = {2022-07-19},
  langid = {english}
}

@article{Neuefeind2006_NanoscaleOrderedMaterials,
  title = {A Nanoscale Ordered Materials Diffractometer for the {{SNS}}},
  author = {Neuefeind, J{\"o}rg and Chipley, Kenneth K. and Tulk, Chris A. and Simonson, J. Michael and Winokur, Michael J.},
  year = 2006,
  month = nov,
  journal = {Physica B: Condensed Matter},
  volume = {385--386},
  pages = {1066--1069},
  issn = {0921-4526},
  doi = {10.1016/j.physb.2006.05.341},
  urldate = {2022-10-25},
  langid = {english}
}

@article{Nguyen2023_ExploringNaClPuCl3Molten,
  title = {Exploring {{NaCl-PuCl3}} Molten Salts with Machine Learning Interatomic Potentials and Graph Theory},
  author = {Nguyen, Manh-Thuong and Glezakou, Vassiliki-Alexandra and Rousseau, Roger and Paviet, Patricia D.},
  year = 2023,
  month = dec,
  journal = {Applied Materials Today},
  volume = {35},
  pages = {101951},
  issn = {2352-9407},
  doi = {10.1016/j.apmt.2023.101951},
  urldate = {2026-08-24}
}

@article{Perdew1996_GeneralizedGradientApproximation,
  title = {Generalized {{Gradient Approximation Made Simple}}},
  author = {Perdew, John P. and Burke, Kieron and Ernzerhof, Matthias},
  year = 1996,
  month = oct,
  journal = {Physical Review Letters},
  volume = {77},
  number = {18},
  pages = {3865--3868},
  publisher = {American Physical Society},
  doi = {10.1103/PhysRevLett.77.3865},
  urldate = {2022-12-02}
}

@article{Peterson2021_IllustratedFormalismsTotal,
  title = {Illustrated Formalisms for Total Scattering Data: {{A}} Guide for New Practitioners},
  author = {Peterson, Peter F. and Olds, Daniel and McDonnell, Marshall T. and Page, Katharine},
  year = 2021,
  journal = {Journal of Applied Crystallography},
  volume = {54},
  pages = {317--332},
  publisher = {International Union of Crystallography},
  issn = {16005767},
  doi = {10.1107/S1600576720015630}
}

@article{Petruska1955_FISSIONYIELDSCESIUM,
  title = {{{THE FISSION YIELDS OF THE CESIUM ISOTOPES FORMED IN THE THERMAL NEUTRON FISSION OF U235 AND THE NEUTRON ABSORPTION CROSS SECTION OF Xe135}}},
  author = {Petruska, J. A. and Melaika, E. A. and Tomlinson, R. H.},
  year = 1955,
  month = nov,
  journal = {Canadian Journal of Physics},
  volume = {33},
  number = {11},
  pages = {640--649},
  publisher = {NRC Research Press},
  issn = {0008-4204},
  doi = {10.1139/p55-080},
  urldate = {2026-05-04}
}

@article{Plimpton1995_FastParallelAlgorithms,
  title = {Fast {{Parallel Algorithms}} for {{Short-Range Molecular Dynamics}}},
  author = {Plimpton, Steve},
  year = 1995,
  month = mar,
  journal = {Journal of Computational Physics},
  volume = {117},
  number = {1},
  pages = {1--19},
  issn = {0021-9991},
  doi = {10.1006/jcph.1995.1039},
  urldate = {2022-12-02},
  langid = {english}
}

@article{Rahman1972_StructureMotionLiquid,
  title = {Structure and {{Motion}} in {{Liquid BeF2}}, {{LiBeF3}}, and {{LiF}} from {{Molecular Dynamics Calculations}}},
  author = {Rahman, A. and Fowler, R. H. and Narten, A. H.},
  year = 1972,
  journal = {The Journal of Chemical Physics},
  volume = {57},
  number = {7},
  pages = {3010--3011},
  issn = {0021-9606},
  doi = {10.1063/1.1678700},
  urldate = {2023-06-21}
}

@article{Rakib2025_StructureMoltenF7LiNaK,
  title = {The Structure of Molten {{F7LiNaK}} with {{CeF3}} Using Neutron Diffraction and {{EPSR}}},
  author = {Rakib, G. S. and Neuefeind, J{\"o}rg C. and Everett, Susan Michelle and Mills, Rebecca and Rose, Melissa A. and Lee, Shao-Chun and Z, Y and Heuser, Brent J.},
  year = 2025,
  month = jun,
  journal = {Journal of Nuclear Materials},
  pages = {156000},
  issn = {0022-3115},
  doi = {10.1016/j.jnucmat.2025.156000},
  urldate = {2025-06-23}
}

@techreport{Riley2018_IdentificationPotentialWaste,
  title = {Identification of {{Potential Waste Processing}} and {{Waste Form Options}} for {{Molten Salt Reactors}}},
  author = {Riley, Brian J. and Mcfarlane, Joanna and DelCul, Guillermo and Vienna, John D. and Contescu, Cristian I. and Hay, Laurie M. and Savino, A. V. and Adkins, Harold E.},
  year = 2018,
  month = aug,
  number = {ORNL/LTR-2018/907},
  institution = {Oak Ridge National Laboratory (ORNL), Oak Ridge, TN (United States)},
  doi = {10.2172/1543229},
  urldate = {2026-04-10},
  langid = {english}
}

@article{Riley2019_MoltenSaltReactor,
  title = {Molten Salt Reactor Waste and Effluent Management Strategies: {{A}} Review},
  shorttitle = {Molten Salt Reactor Waste and Effluent Management Strategies},
  author = {Riley, Brian J. and McFarlane, Joanna and DelCul, Guillermo D. and Vienna, John D. and Contescu, Cristian I. and Forsberg, Charles W.},
  year = 2019,
  month = apr,
  journal = {Nuclear Engineering and Design},
  volume = {345},
  pages = {94--109},
  issn = {0029-5493},
  doi = {10.1016/j.nucengdes.2019.02.002},
  urldate = {2026-04-10}
}

@techreport{Robertson1965_MSREDesignOperations,
  title = {{{MSRE Design}} \& {{Operations Report Part}} 1 {{Description}} of {{Reactor Design}}},
  author = {Robertson, R. C.},
  year = 1965,
  month = jan,
  number = {ORNL-TM-728},
  institution = {Oak Ridge National Lab. (ORNL), Oak Ridge, TN (United States)},
  doi = {10.2172/4654707},
  urldate = {2023-03-15},
  langid = {english}
}

@article{Rodriguez2021_ThermodynamicTransportProperties,
  title = {Thermodynamic and {{Transport Properties}} of {{LiF}} and {{FLiBe Molten Salts}} with {{Deep Learning Potentials}}},
  author = {Rodriguez, Alejandro and Lam, Stephen and Hu, Ming},
  year = 2021,
  month = nov,
  journal = {ACS Applied Materials \& Interfaces},
  volume = {13},
  number = {46},
  pages = {55367--55379},
  issn = {1944-8244},
  doi = {10.1021/acsami.1c17942},
  urldate = {2026-08-28}
}

@article{Rollet2011_StudiesLocalStructures,
  title = {Studies of the Local Structures of Molten Metal Halides},
  author = {Rollet, Anne Laure and Salanne, Mathieu},
  year = 2011,
  journal = {Annual Reports on the Progress of Chemistry - Section C},
  volume = {107},
  pages = {88--123},
  issn = {02601826},
  doi = {10.1039/c1pc90003j}
}

@article{Roper2022_MoltenSaltAdvanced,
  title = {Molten Salt for Advanced Energy Applications: {{A}} Review},
  shorttitle = {Molten Salt for Advanced Energy Applications},
  author = {Roper, Robin and Harkema, Megan and Sabharwall, Piyush and Riddle, Catherine and Chisholm, Brandon and Day, Brandon and Marotta, Paul},
  year = 2022,
  month = may,
  journal = {Annals of Nuclear Energy},
  volume = {169},
  pages = {108924},
  issn = {0306-4549},
  doi = {10.1016/j.anucene.2021.108924},
  urldate = {2026-01-06}
}

@article{Rosenthal1970_MoltenSaltReactorsHistory,
  title = {Molten-{{Salt Reactors}}---{{History}}, {{Status}}, and {{Potential}}},
  author = {Rosenthal, M. W. and Kasten, P. R. and Briggs, R. B.},
  year = 1970,
  month = feb,
  journal = {Nuclear Applications and Technology},
  volume = {8},
  number = {2},
  pages = {107--117},
  issn = {0550-3043},
  doi = {10.13182/NT70-A28619},
  urldate = {2023-02-14}
}

@article{Salanne2006_FirstprinciplesDescriptionLiquid,
  title = {A First-Principles Description of Liquid {{BeF2}} and Its Mixtures with {{LiF}}: 2. {{Network}} Formation in {{LiF-BeF2}}},
  author = {Salanne, Mathieu and Simon, Christian and Turq, Pierre and Heaton, Robert J. and Madden, Paul A.},
  year = 2006,
  journal = {Journal of Physical Chemistry B},
  volume = {110},
  number = {23},
  pages = {11461--11467},
  issn = {15206106},
  doi = {10.1021/jp061002u},
  pmid = {16771420}
}

@article{Serp2014_MoltenSaltReactor,
  title = {The Molten Salt Reactor ({{MSR}}) in Generation {{IV}}: {{Overview}} and Perspectives},
  author = {Serp, J{\'e}r{\^o}me and Allibert, Michel and Bene{\v s}, Ond{\v r}ej and Delpech, Sylvie and Feynberg, Olga and Ghetta, V{\'e}ronique and Heuer, Daniel and Holcomb, David and Ignatiev, Victor and Kloosterman, Jan Leen and Luzzi, Lelio and {Merle-Lucotte}, Elsa and Uhl{\'i}{\v r}, Jan and Yoshioka, Ritsuo and Zhimin, Dai},
  year = 2014,
  journal = {Progress in Nuclear Energy},
  volume = {77},
  pages = {308--319},
  issn = {01491970},
  doi = {10.1016/j.pnucene.2014.02.014}
}

@article{Smith2020_NewApproachCoupled,
  title = {A New Approach for Coupled Modelling of the Structural and Thermo-Physical Properties of Molten Salts. {{Case}} of a Polymeric Liquid {{LiF-BeF2}}},
  author = {Smith, A. L. and Capelli, E. and Konings, R. J. M. and Gheribi, A. E.},
  year = 2020,
  month = feb,
  journal = {Journal of Molecular Liquids},
  volume = {299},
  pages = {112165},
  issn = {0167-7322},
  doi = {10.1016/j.molliq.2019.112165},
  urldate = {2023-03-28},
  langid = {english}
}

@article{Soper1996_EmpiricalPotentialMonte,
  title = {Empirical Potential {{Monte Carlo}} Simulation of Fluid Structure},
  author = {Soper, A. K.},
  year = 1996,
  month = jan,
  journal = {Chemical Physics},
  volume = {202},
  number = {2},
  pages = {295--306},
  issn = {0301-0104},
  doi = {10.1016/0301-0104(95)00357-6},
  urldate = {2025-11-26}
}

@article{Soper2001_TestsEmpiricalPotential,
  title = {Tests of the Empirical Potential Structure Refinement Method and a New Method of Application to Neutron Diffraction Data on Water},
  author = {Soper, A. K.},
  year = 2001,
  month = sep,
  journal = {Molecular Physics},
  volume = {99},
  number = {17},
  pages = {1503--1516},
  publisher = {Taylor \& Francis},
  issn = {0026-8976},
  doi = {10.1080/00268970110056889},
  urldate = {2025-11-26}
}

@article{Soper2005_PartialStructureFactors,
  title = {Partial Structure Factors from Disordered Materials Diffraction Data: {{An}} Approach Using Empirical Potential Structure Refinement},
  shorttitle = {Partial Structure Factors from Disordered Materials Diffraction Data},
  author = {Soper, A. K.},
  year = 2005,
  month = sep,
  journal = {Physical Review B},
  volume = {72},
  number = {10},
  pages = {104204},
  publisher = {American Physical Society},
  doi = {10.1103/PhysRevB.72.104204},
  urldate = {2025-11-24}
}

@article{Sorbom2015_ARCCompactHighfield,
  title = {{{ARC}}: {{A}} Compact, High-Field, Fusion Nuclear Science Facility and Demonstration Power Plant with Demountable Magnets},
  shorttitle = {{{ARC}}},
  author = {Sorbom, B. N. and Ball, J. and Palmer, T. R. and Mangiarotti, F. J. and Sierchio, J. M. and Bonoli, P. and Kasten, C. and Sutherland, D. A. and Barnard, H. S. and Haakonsen, C. B. and Goh, J. and Sung, C. and Whyte, D. G.},
  year = 2015,
  month = nov,
  journal = {Fusion Engineering and Design},
  volume = {100},
  pages = {378--405},
  issn = {0920-3796},
  doi = {10.1016/j.fusengdes.2015.07.008},
  urldate = {2023-03-27},
  langid = {english}
}

@article{Sprouster2022_MolecularStructurePhasea,
  title = {Molecular {{Structure}} and {{Phase Equilibria}} of {{Molten Fluoride Salt}} with and without {{Dissolved Cesium}}: {{FLiNaK}}--{{CsF}} (5 Mol \%)},
  shorttitle = {Molecular {{Structure}} and {{Phase Equilibria}} of {{Molten Fluoride Salt}} with and without {{Dissolved Cesium}}},
  author = {Sprouster, David and Zheng, Guiqiu and Lee, Shao-Chun and Olds, Daniel and Agca, Can and McFarlane, Joanna and Z, Y and Khaykovich, Boris},
  year = 2022,
  month = jul,
  journal = {ACS Applied Energy Materials},
  volume = {5},
  number = {7},
  pages = {8067--8074},
  publisher = {American Chemical Society},
  doi = {10.1021/acsaem.2c00544},
  urldate = {2023-08-08}
}

@incollection{Sridharan2013_12CorrosionMolten,
  title = {12 - {{Corrosion}} in {{Molten Salts}}},
  booktitle = {Molten {{Salts Chemistry}}},
  author = {Sridharan, K. and Allen, T. R.},
  editor = {Lantelme, Fr{\'e}d{\'e}ric and Groult, Henri},
  year = 2013,
  month = jan,
  pages = {241--267},
  publisher = {Elsevier},
  address = {Oxford},
  doi = {10.1016/B978-0-12-398538-5.00012-3},
  urldate = {2022-07-13},
  isbn = {978-0-12-398538-5},
  langid = {english}
}

@article{Vaslow1973_DiffractionPatternStructure,
  title = {Diffraction Pattern and Structure of Molten {{BeF2-LiF}} Solutions},
  author = {Vaslow, F. and Narten, A. H.},
  year = 1973,
  journal = {The Journal of Chemical Physics},
  volume = {59},
  number = {9},
  pages = {4955--4960},
  issn = {00219606},
  doi = {10.1063/1.1680711}
}

@article{Volkovich2003_TreatmentMoltenSalt,
  title = {Treatment of Molten Salt Wastes by Phosphate Precipitation: Removal of Fission Product Elements after Pyrochemical Reprocessing of Spent Nuclear Fuels in Chloride Melts},
  shorttitle = {Treatment of Molten Salt Wastes by Phosphate Precipitation},
  author = {Volkovich, Vladimir A and Griffiths, Trevor R and Thied, Robert C},
  year = 2003,
  month = nov,
  journal = {Journal of Nuclear Materials},
  volume = {323},
  number = {1},
  pages = {49--56},
  issn = {0022-3115},
  doi = {10.1016/j.jnucmat.2003.08.024},
  urldate = {2026-04-28}
}

@article{Wang2018_DeePMDkitDeepLearning,
  title = {{{DeePMD-kit}}: {{A}} Deep Learning Package for Many-Body Potential Energy Representation and Molecular Dynamics},
  shorttitle = {{{DeePMD-kit}}},
  author = {Wang, Han and Zhang, Linfeng and Han, Jiequn and E, Weinan},
  year = 2018,
  month = jul,
  journal = {Computer Physics Communications},
  volume = {228},
  pages = {178--184},
  issn = {0010-4655},
  doi = {10.1016/j.cpc.2018.03.016},
  urldate = {2022-12-02},
  langid = {english}
}

@article{Wang2021_StructuresThoriumFluoride,
  title = {On the {{Structures}} of {{Thorium Fluoride}} and {{Oxyfluoride Anions}} in {{Molten FLiBe}} and {{FLiNaK}}},
  author = {Wang, Chenyang and Chen, Xiuting and Gong, Yu},
  year = 2021,
  month = feb,
  journal = {The Journal of Physical Chemistry B},
  volume = {125},
  number = {6},
  pages = {1640--1646},
  publisher = {American Chemical Society},
  issn = {1520-6106},
  doi = {10.1021/acs.jpcb.0c10197},
  urldate = {2023-04-03}
}

@article{Wang2022_FirstprinciplesMolecularDynamics,
  title = {First-Principles Molecular Dynamics Study of the Behavior of Tritium in Molten {{LiF-BeF2}} Eutectic},
  author = {Wang, Hui and Yue, Baohua and Yan, Liuming and Jiang, Tao and Peng, Shuming},
  year = 2022,
  month = jan,
  journal = {Journal of Molecular Liquids},
  volume = {345},
  pages = {117027},
  issn = {0167-7322},
  doi = {10.1016/j.molliq.2021.117027},
  urldate = {2023-03-16},
  langid = {english}
}

@article{Wang2025_CoordinationdrivenMixingBehavior,
  title = {Coordination-Driven Mixing Behavior of Corrosion and Fission Products in {{NaCl-UCl}}{$_3$} Molten Salt},
  author = {Wang, Gaoxue and Li, Bo and Yang, Ping and Besmann, Theodore M. and Andersson, David A.},
  year = 2025,
  month = sep,
  journal = {Journal of Nuclear Materials},
  volume = {615},
  pages = {155949},
  issn = {0022-3115},
  doi = {10.1016/j.jnucmat.2025.155949},
  urldate = {2026-08-24}
}

@phdthesis{Williams_StructureSpeciationMolten,
  title = {Structure and {{Speciation}} in {{Molten Fluoride Salts}} and {{Knowing}} and {{Being}} as a {{Nuclear Engineer}}},
  author = {Williams, Haley},
  year = 2025,
  langid = {english}
}

@article{Williams2008_EvaluationSaltCoolants,
  title = {Evaluation of {{Salt Coolants}} for {{Reactor Applications}}},
  author = {Williams, D. F. and Clarno, K. T.},
  year = 2008,
  month = sep,
  journal = {Nuclear Technology},
  volume = {163},
  number = {3},
  pages = {330--343},
  publisher = {Taylor \& Francis},
  issn = {0029-5450},
  doi = {10.13182/NT08-A3992},
  urldate = {2022-08-17}
}

@techreport{Williams2017_TechnologyAppliedRD,
  type = {Technical {{Report}}},
  title = {Technology and {{Applied R}}\&{{D Needs}} for {{Molten Salt Chemistry}}},
  author = {Williams, D F and Britt, P F},
  year = 2017,
  address = {Oak Ridge, TN},
  institution = {Oak Ridge National Laboratory (ORNL)}
}

@article{Wilson2016_StructureDynamicsNetworkforming,
  title = {Structure and Dynamics in Network-Forming Materials},
  author = {Wilson, Mark},
  year = 2016,
  month = oct,
  journal = {Journal of Physics: Condensed Matter},
  volume = {28},
  number = {50},
  pages = {503001},
  publisher = {IOP Publishing},
  issn = {0953-8984},
  doi = {10.1088/0953-8984/28/50/503001},
  urldate = {2022-12-22},
  langid = {english}
}

@article{Winner2021_AbinitioSimulationStudies,
  title = {Ab-Initio Simulation Studies of Chromium Solvation in Molten Fluoride Salts},
  author = {Winner, Nicholas and Williams, Haley and Scarlat, Raluca O. and Asta, Mark},
  year = 2021,
  journal = {Journal of Molecular Liquids},
  volume = {335},
  pages = {116351},
  publisher = {The Authors},
  issn = {01677322},
  doi = {10.1016/j.molliq.2021.116351}
}

@article{Wright1989_NeutronDiffractionMolecular,
  title = {A Neutron Diffraction and Molecular Dynamics Investigation of the Structure of Vitreous Beryllium Fluoride},
  author = {Wright, Adrian C. and Clare, Alexis G. and Etherington, George and Sinclair, Roger N. and Brawer, Steven A. and Weber, Marvin J.},
  year = 1989,
  month = nov,
  journal = {Journal of Non-Crystalline Solids},
  volume = {111},
  number = {2},
  pages = {139--152},
  issn = {0022-3093},
  doi = {10.1016/0022-3093(89)90275-5},
  urldate = {2023-05-08},
  langid = {english}
}

@article{Wright1994_NeutronScatteringVitreous,
  title = {Neutron Scattering from Vitreous Silica. {{V}}. {{The}} Structure of Vitreous Silica: {{What}} Have We Learned from 60 Years of Diffraction Studies?},
  shorttitle = {Neutron Scattering from Vitreous Silica. {{V}}. {{The}} Structure of Vitreous Silica},
  author = {Wright, Adrian C.},
  year = 1994,
  month = nov,
  journal = {Journal of Non-Crystalline Solids},
  series = {Proceedings of the {{First PAC RIM Meeting}} on {{Glass}} and {{Optical Materials}}},
  volume = {179},
  pages = {84--115},
  issn = {0022-3093},
  doi = {10.1016/0022-3093(94)90687-4},
  urldate = {2023-06-08},
  langid = {english}
}

@article{Yin2025_LocalStructureIonic,
  title = {Local {{Structure}} and {{Ionic Diffusion}} in {{LiF-BeF2-ThF4 Molten Salts}}: {{Insights}} from {{Ab Initio Molecular Dynamics}}},
  shorttitle = {Local {{Structure}} and {{Ionic Diffusion}} in {{LiF-BeF2-ThF4 Molten Salts}}},
  author = {Yin, Yuan and Liang, Wenshuo and Wang, Dezhong and Zhou, Wentao},
  year = 2025,
  month = apr,
  journal = {The Journal of Physical Chemistry B},
  publisher = {American Chemical Society},
  issn = {1520-6106},
  doi = {10.1021/acs.jpcb.5c01173},
  urldate = {2025-04-24}
}

@article{Zaghloul2003_ThermoPhysicalPropertiesEquilibrium,
  title = {Thermo-{{Physical Properties}} and {{Equilibrium Vapor-Composition}} of {{Lithium Fluoride-Beryllium Fluoride}} ({{2LiF}}/{{BeF2}}) {{Molten Salt}}},
  author = {Zaghloul, Mofreh R. and Sze, Dai Kai and Raffray, A. Ren{\'e}},
  year = 2003,
  month = sep,
  journal = {Fusion Science and Technology},
  volume = {44},
  number = {2},
  pages = {344--350},
  publisher = {Taylor \& Francis},
  issn = {1536-1055},
  doi = {10.13182/FST03-A358},
  urldate = {2022-12-02}
}

@article{Zakiryanov2026_LiFBeF2MoltenMixtures,
  title = {{{LiF-BeF2}} Molten Mixtures Studied Using a Composition-Transferable Deep Learning Potential},
  author = {Zakiryanov, Dmitry},
  year = 2026,
  month = jun,
  journal = {Computational Materials Science},
  volume = {270},
  pages = {114775},
  issn = {09270256},
  doi = {10.1016/j.commatsci.2026.114775},
  urldate = {2026-06-02},
  langid = {english}
}

@inproceedings{Zhang2018_EndtoendSymmetryPreserving,
  title = {End-to-End {{Symmetry Preserving Inter-atomic Potential Energy Model}} for {{Finite}} and {{Extended Systems}}},
  booktitle = {Advances in {{Neural Information Processing Systems}}},
  author = {Zhang, Linfeng and Han, Jiequn and Wang, Han and Saidi, Wissam and Car, Roberto and E, Weinan},
  year = 2018,
  volume = {31},
  publisher = {Curran Associates, Inc.},
  address = {Montr\'eal},
  urldate = {2022-12-02}
}

@article{Zhou2021_ExperimentalMethodQuantify,
  title = {Experimental Method to Quantify the Ring Size Distribution in Silicate Glasses and Simulation Validation Thereof},
  author = {Zhou, Qi and Shi, Ying and Deng, Binghui and Neuefeind, J{\"o}rg and Bauchy, Mathieu},
  year = 2021,
  month = jul,
  journal = {Science Advances},
  volume = {7},
  number = {28},
  pages = {eabh1761},
  issn = {2375-2548},
  doi = {10.1126/sciadv.abh1761},
  urldate = {2023-04-27},
  langid = {english}
}

@book{Zhou2022_MolecularDynamicsSimulation,
  title = {Molecular Dynamics Simulation: Fundamentals and Applications},
  shorttitle = {Molecular Dynamics Simulation},
  author = {Zhou, Kun and Liu, Bo},
  year = 2022,
  publisher = {Elsevier},
  address = {Amsterdam Kidlington Cambridge, Mass},
  isbn = {978-0-12-816419-8},
  langid = {english}
}

\end{document}